\documentclass[pre,twocolumn,showpacs,superscriptaddress]{revtex4-1}
\usepackage{amsmath}
\usepackage[xdvi]{graphicx}%
\usepackage{dcolumn}
\usepackage{amsmath}
\usepackage{color}
\usepackage{bm}
\usepackage{subfigure}

\usepackage{color}
\definecolor{darkblue}{rgb}{0,0,0.6}
\definecolor{darkred}{rgb}{0.6,0,0}
\definecolor{darkgreen}{rgb}{0,0.6,0}

\usepackage[colorlinks=true,urlcolor=darkblue,citecolor=darkblue,linkcolor=darkred,hyperfootnotes=false]{hyperref}

\DeclareMathOperator*{\argmax}{arg\,max}

\newcommand{\dd}{\mathrm{d}}
\newcommand{\ee}{\mathrm{e}}

\newcommand{\xt}{{\tilde x}}

\begin{document}

\title{
Population dynamics method with a multi-canonical feedback control}

\author{Takahiro Nemoto}
\affiliation{Laboratoire de Probabilit\'es et Mod\`eles Al\'eatoires, Sorbonne Paris Cit\'e, UMR 7599 CNRS, Universit\'e Paris Diderot, 75013 Paris, France}
\affiliation{Laboratoire de Physique, ENS de Lyon, Universit\'e de Lyon, CNRS, 46 all\'ee d'Italie, 69364 Lyon, France}
\author{Freddy Bouchet}
\affiliation{Laboratoire de Physique, ENS de Lyon, Universit\'e de Lyon, CNRS, 46 all\'ee d'Italie, 69364 Lyon, France}
\author{Robert L. Jack}
\affiliation{Department of Physics, University of Bath, Bath BA2 7AY, United Kingdom}
\author{Vivien Lecomte}
\affiliation{Laboratoire de Probabilit\'es et Mod\`eles Al\'eatoires, Sorbonne Paris Cit\'e, UMR 7599 CNRS, Universit\'e Paris Diderot, 75013 Paris, France}

\date{\today}

\begin{abstract}
We discuss the Giardin\`a-Kurchan-Peliti population dynamics method for evaluating large deviations of time averaged quantities in Markov processes [Phys. Rev. Lett. \textbf{96}, 120603 (2006)].  This method exhibits systematic errors which can be large in some circumstances, particularly for systems with weak noise, with many degrees of freedom,  or close to dynamical phase transitions.  We show how these errors can be mitigated by introducing control forces within the algorithm.  These forces are determined by an iteration-and-feedback scheme, inspired by multicanonical methods in equilibrium sampling.  We demonstrate substantially improved results in a simple model and we discuss potential applications to more complex systems.
\end{abstract}


\maketitle

\section{Introduction}
In many physical systems, interesting and important behaviour is associated with rare events -- examples include crystal nucleation, slow transitions in biomolecules~\cite{auer01,sear07,ren05}, rare transitions in turbulent flows~\cite{VKS,BouchetSimonnet}, and extreme events in climate dynamics~\cite{ClimateExtremes}.  Many computational methods for sampling these events have been proposed and exploited~\cite{auer01,ren05,BouchetSimonnet,Chandler,FFS1,Splittingmethod,Populationdynamics,TailleurKurchan, Berg_Neuhaus1, Wang_Landau1, OrtizLaelbling, DupuisWang, CappeDoucGuillin, ChanLai, Nemoto_Sasa_PRL}.  One family of methods is based around population dynamics~\cite{DMC,Aldous,Grassberger,Lelievre,Garnier,Rolland}, in which several copies of a system evolve in parallel: the copies which exhibit the rare behaviour of interest are copied (or \emph{cloned}) while other copies are discarded.  The result is that \emph{typical} copies within the population dynamics reproduce the desired rare events in the original system.  One such method has recently been employed to characterise a particular class of rare events~\cite{Populationdynamics,TailleurKurchan}, in which time-averaged physical quantities exhibit  \emph{large deviations}~\cite{Touchette,Dembo_Zeitouni} from their typical values in the large time limit.
Studies of such events have revealed new and unexpected features in
glass-formers~\cite{Hedges}, biomolecules~\cite{picciani11,weber13,mey14}, non-equilibrium transport~\cite{derrida07,hurtado14} and integrable systems~\cite{TailleurKurchan}. 
In this article, we identify a pitfall that limits the computational efficiency of the population dynamics method, and we show that the method can be modified so as to avoid this problem. 
The issue at stake is the number of copies of the system that must be considered in order to obtain accurate results -- if very many copies are required then the method is difficult to apply, especially if even a single system is complex or contains many degrees of freedom.  In some relevant cases then the standard population dynamics method requires an exponentially large population to be effective~\cite{hurtado09}.  However, the method that we propose here, which is an improved version of the population dynamics, inspired by multicanonical methods in equilibrium systems~\cite{Berg_Neuhaus1,Wang_Landau1} (or adaptive importance sampling \cite{OrtizLaelbling,DupuisWang,CappeDoucGuillin,ChanLai}), can still be effective in these cases. 

The intuitive description of the problem that we identify is the following. The population dynamics is characterised by two different distributions, which describe the state of the system at some fixed final time, and its state at intermediate times. We show that in situations where the two distributions have a small overlap, the population dynamics is affected by a serious sampling problem, in which statistical estimators of the quantities of interest become dominated by just a few samples.  One relevant case is that of systems with weak noise, for which the two distributions become more and more concentrated around their most likely values, so that they quite generally have zero overlap: this  leads to an unavoidable failure of the population dynamics.
In this article, we describe how to modify the population dynamics so as to maintain the two distributions close to each other, thus solving the sampling problem.
We argue that this new method will provide a step-change in the complexity of the systems for which large deviation computations can be performed.

The structure of the paper is as follows: 
we introduce our model and the population dynamics algorithm in Section~\ref{Sec:SetUp}.
We discuss sampling problems associated with this algorithm in Section~\ref{Sec:SamplingErrors}. 
In Section~\ref{Sec:FB_PD}, we introduce our main idea, which is to combine a controlling force with the population dynamics algorithm, in order to resolve the sampling issues. In Section~\ref{Sec:NumEx}, we numerically demonstrate this method in a simple Brownian particle model. Finally, in Section~\ref{Sec:outlook}, we describe the potential for future applications and extensions of this work.

\section{Model and Methods}
\label{Sec:SetUp}

\subsection{ Rare event problem}
The rare events that we consider can take place in a variety of models.  To illustrate the method,  
consider a particle moving in $d$-dimensions, whose position $x\in {\bf R}^{d}$ obeys a Langevin equation
\begin{equation}
\dot{x}_t = F(x_t) + B(x_t) \xi_t,
 \label{equ:dx}
\end{equation}
where $\xi$ is a $d$-dimensional Gaussian white noise of unit variance,
$F(x)\in {\bf R}^{d}$ a deterministic force, and the matrix $B(x)$ specifies the action of the noise on the particle.  %$B(x)\equiv \sqrt{\epsilon \sigma(x)} $ with a $d\times d$ matrix $\sigma(x)$ and a noise intensity $\epsilon$, which specify the action of the noise on the particle %
~\cite{footnoteinteraction}.
We use the It\=o convention \cite{Gardiner} for stochastic calculus throughout this paper, although one can also work with the Stratonovich convention by using a transformation formula to relate one convention to the other~\cite{FootnoteConvention_Multiplication}.

We restrict to ergodic systems, and we focus on rare events in which a time averaged quantity $\Lambda(\tau)$ 
takes some non-typical value.  Here $\tau$ is the long time period over which the average is taken, and
\begin{equation}
 \Lambda(\tau) = \Lambda_{\rm d}(\tau) + \Lambda_{\rm c}(\tau)
\label{Observe_general}
\end{equation}
consists of a ``scalar'' contribution 
\begin{equation}
\Lambda_{\rm d}(\tau) =\frac{1}{\tau} \int_{0}^{\tau} \lambda_{\rm d}(x_t) \dd t
\label{Observe_Scalar}
\end{equation}
and a ``vector'' one
\begin{equation}
\Lambda_{\rm c}(\tau) = \frac{1}{\tau} \int_{0}^{\tau}\lambda_{\rm c}(x_t) \cdot  \dd x_t ,
\label{Observe_Vector}
\end{equation}
 where $\lambda_{{\rm d},{\rm c}}$ are arbitrary functions of the particle position $x$. The first contribution $\Lambda_{\rm d}(\tau)$ is a time-average of a quantity $\lambda_{\rm d}$ that depends only on the position $x$ (i.e. a time-average of a static function such as a particle density or an energy density), whereas the second contribution $\Lambda_{\rm c}(\tau)$ includes transitions of $x$ as seen from the form $\lambda_{\rm c}(x_t)\cdot \dd x_t$ (i.e.~$\Lambda_{\rm c}(\tau)$ is an average of a dynamic function such as a particle current or an energy current \cite{Footnote_Work}). See also the explanation around eq.(34) in \cite{Chetrite_Touchette2} for a pedagogical introduction of $\Lambda(\tau)$.
This class of observable includes many physically and mathematically interesting examples, and fluctuations of these quantities have been intensively studied recently, where examples are entropy production~\cite{lebowitz99,lecomte07jsp}, dynamical activity~\cite{garrahan07,Hedges}, and particle fluxes~\cite{bodineau05}.

In the limit of large $\tau$, ergodicity of the system means that the observable $\Lambda(\tau)$ is almost surely equal to its
typical value $\overline\Lambda$.  Our aims are (i) to estimate the (small) probability of deviations
from this value, and (ii) to generate the rare trajectories that lead to these deviations.  This is an important problem because these
non-typical trajectories can exhibit interesting and unusual structures, including misfolded proteins~\cite{weber13,mey14}, stable glass states~\cite{Hedges} 
and travelling waves in models of particle transport~\cite{hurtado14}.

To achieve these aims, the standard theoretical route~\cite{lebowitz99,Touchette} is to introduce a \emph{biasing field} $h$, which controls 
 deviations of $\Lambda(\tau)$ from its typical value.  Specifically, we consider an ensemble of paths $X=(x_t)_{t=0}^\tau$ with (unnormalised) probability density 
\begin{equation}
P_h[X] = \pi_0(x_0) \exp\left[ -\int_0^\tau {\cal L}(x_t,\dot{x}_t) \dd t + h \tau\Lambda(\tau) \right], 
\end{equation}
where
\begin{equation}
{\cal L}(x,\dot x) = \tfrac{1}{2} [\dot{x} - F(x)] \cdot \kappa(x)^{-1} [\dot{x} - F(x)]
\label{eq:Lagrangian}
\end{equation}
is a Lagrangian density that describes the (unbiased) model (\ref{equ:dx}); 
$\pi_0(x)$ is the initial condition for the trajectories, that can be arbitrary and which we take to be the stationary probability distribution of the unbiased model in the numerical examples.
Also, $\kappa=B B^T$ where the notation $B^T$ indicates a matrix transpose~\cite{footnoteinvertible}.

Normalised averages with respect to $P_h$ are denoted by $\langle \cdot \rangle_h$ and we use these averages 
to characterise the rare trajectories associated with deviations of $\Lambda(\tau)$ from $\overline{\Lambda}$, for the model in Eq.~\eqref{equ:dx}.
We define the scaled cumulant generating function (CGF):
\begin{equation}
G(h) = \lim_{\tau \rightarrow \infty} \tau^{-1} \log  \langle  \ee^{ \tau h \Lambda (\tau)} \rangle_0.
\end{equation}
%\label{cumlantgen_epsilon}
In the limit of large $\tau$, the probability distribution of $\Lambda(\tau)$ satisfies a large deviation principle, and can
be obtained by a Legendre transformation of $G(h)$, (for which we assume that the large deviation function of $\Lambda(\tau)$ is convex ~\cite{Touchette,Dembo_Zeitouni}.) In the same limit, for a given deviation $\Lambda$ from $\overline{\Lambda}$, there exists a bias $h^\star(\Lambda)$ for which 
$\langle \cdot \rangle_{h^\star(\Lambda)}$ is equivalent to a 
conditional average over trajectories with $\Lambda(\tau) = \Lambda$ \cite{HTouchette_equivalence}. 
Biased averages with respect to the biased distribution $P_h$, which are numerically evaluated through the population dynamics, thus enable to characterise the trajectories of the original dynamics for which time-averaged physical quantities exhibit large deviations from their typical values in the large time limit.

\subsection{Population dynamics method}
\label{Subsec:PopulationDynamics}
There are several computational methods that allow evaluation of averages with respect to $P_h$~\cite{Populationdynamics,Populationdynamics3,Chandler,Nemoto_Sasa_PRL}.
In the population dynamics method~\cite{Populationdynamics}, one considers $N_c$ copies (or clones) of the system.  
These clones evolve independently as a function of the time $t$, except that (for $h>0$) clones with small $\Lambda(t)$ are periodically removed (eliminated) from the system, while clones with large $\Lambda(t)$ are duplicated (cloned),
to maintain a constant population.
The algorithm is illustrated in Fig.~\ref{fig:traj} and described fully in Appendix~\ref{PDSection_Algorithm}.
This method biases the dynamics towards the rare events of interest.  
For sufficiently large $N_c$ (and large enough $\tau$), 
the method provides accurate estimates of $G(h)$ and it generates sample paths consistent with the biased distribution $P_h$.

%%%%%%%
\begin{figure}
\centering
%\subfigure[~Trajectories $x^a(t)$ generated by population dynamics]{
%\hspace*{-2mm}
\includegraphics[width=0.82\columnwidth]{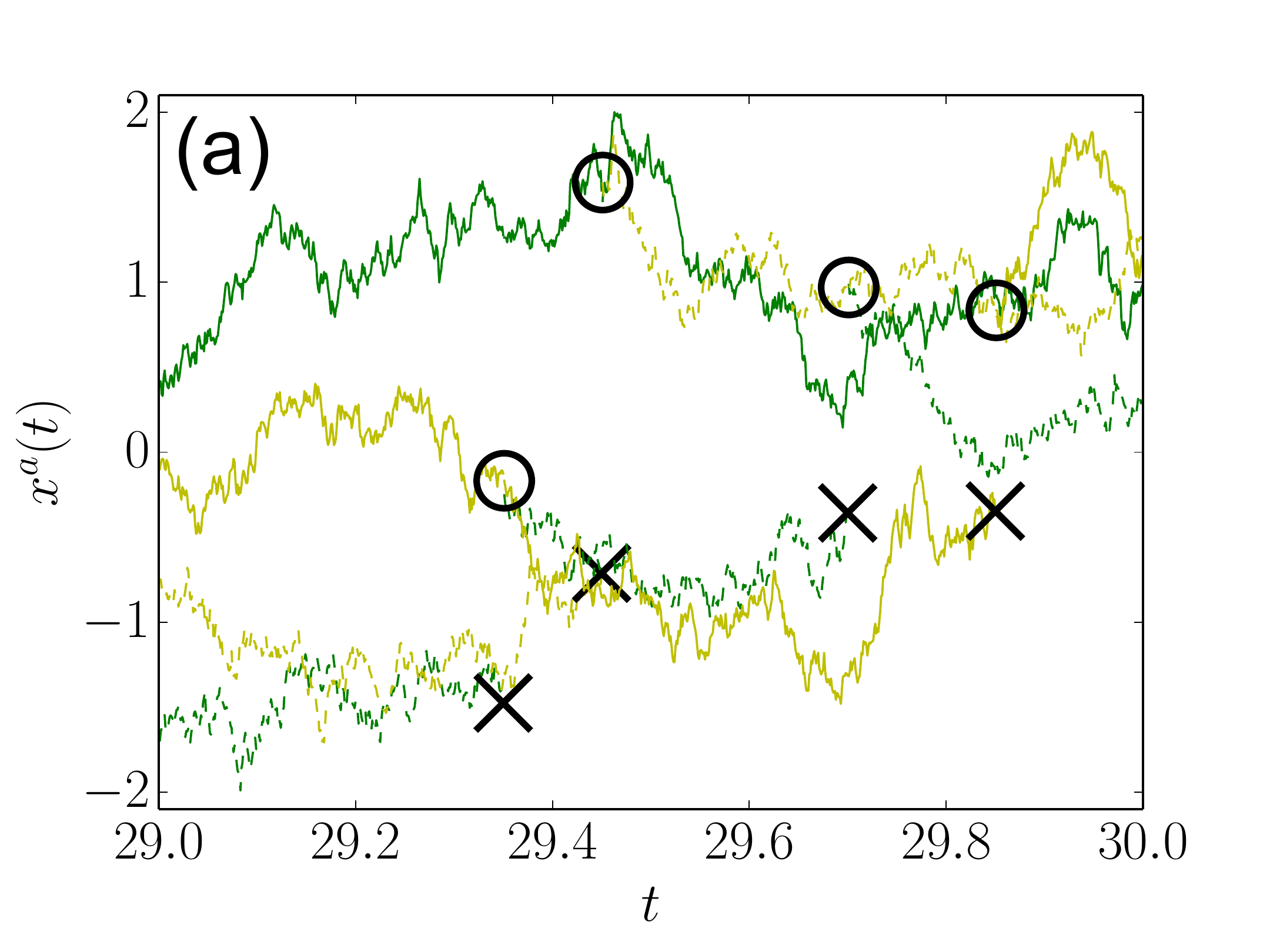}%}
\\
%\subfigure[~Trajectories $\tilde x^a(t)$ for $P_h$]{
%\hspace*{-3mm}
\includegraphics[width=0.82\columnwidth]{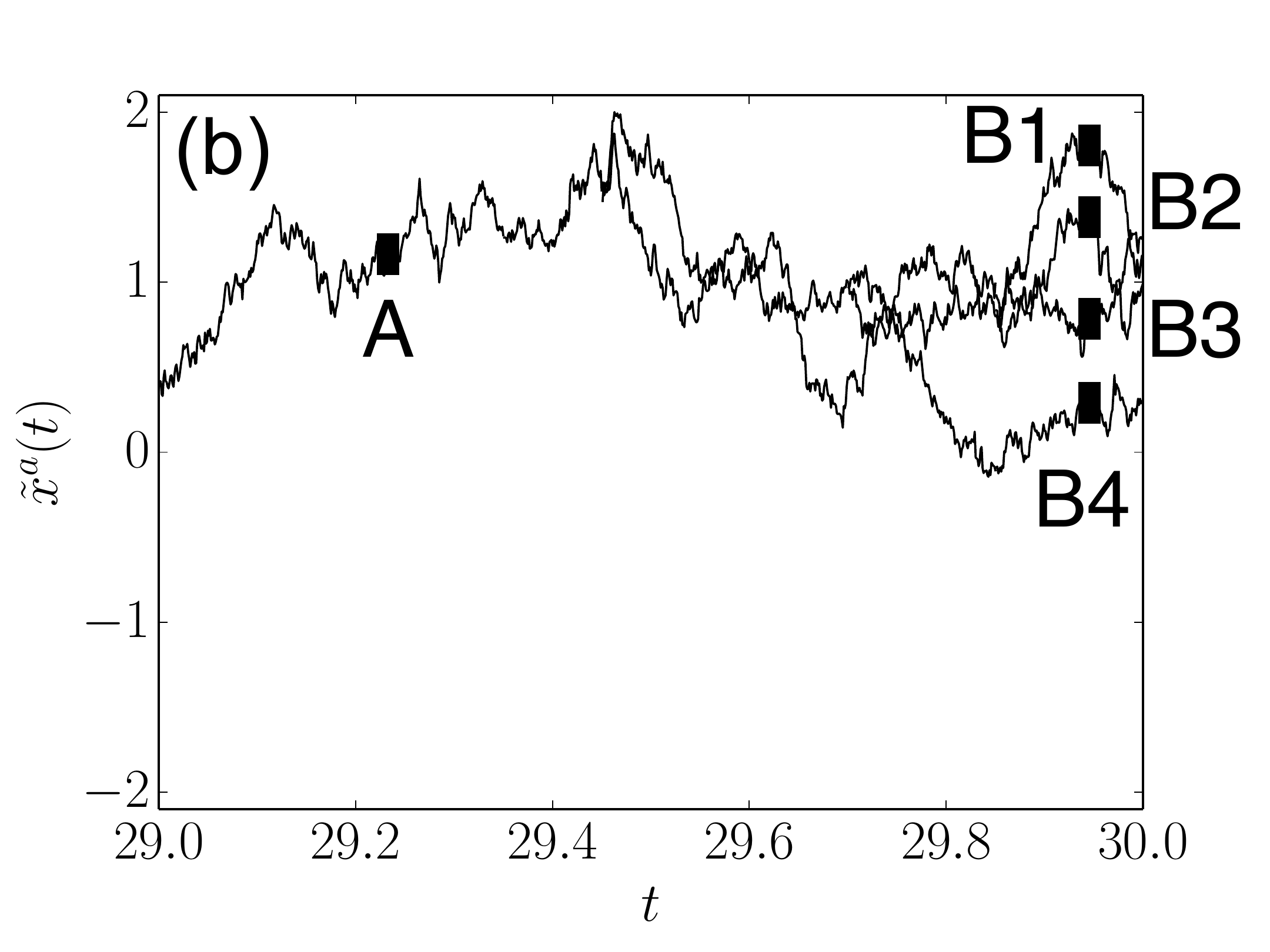}%}
\caption{ \label{Wholetrajectoryonedx_m1k0a1b1hm1dd01N100Step01} (Color online)
\textbf{(a)} Trajectories $x^a(t)$ generated by population dynamics at fixed total population~$N_c=4$ for the model system described in 
Section~\ref{Subsec:NumericalExample} ($\epsilon=1,h=1$). The different colours or line types, which are the green (dark grey in the printed version) solid line, green dashed line, yellow (light grey in the printed version) solid line and yellow dashed line, represent different copies.
%There are $N_c=4$ clones of the system.
At certain times, some copies of the system are removed ($\times$ symbols) and others are duplicated~($\circ$ symbols). The time interval $\Delta T$ for the 
cloning procedure is set to be 0.05, and the time step for solving the Langevin equation is 0.001 (See Appendix~\ref{PDSection_Algorithm} for the details of the algorithm). 
\textbf{(b)} Representative sample paths $\tilde x^a(t)$ for the distribution $P_h[X]$, derived from those in (a) by keeping only trajectories surviving up to final time $\tau=30$. For each cloning event, we also copy the history of the trajectory, which replaces the history of the eliminated trajectory.  This means that the trajectories $(\tilde x^a(t))_{a=1}^{N_c}$ overlap, especially for early times. For example, in panel (b), the point A appears in the past of the four points B1,$\dots$,B4.  For any point $x^a(t)$ (such as A,B1,B2,$\dots$), we define the multiplicity $m_a(t,\tau)$ as the number of trajectories that include this point, and survive until the final time $\tau$.  So for point $A$, the multiplicity is $m_a(t,\tau)=4$ but for B1,$\dots$,B4 then $m_a(t,\tau)=1$.
(For all points in the trajectories who die before $\tau$, which are not drawn in the panel (b), $m_a(t,\tau)=0$.)
%At a given time $t$, we call the number of overlaps
%the multiplicity of a clone $a$, and denote it by $m_a(t)$: e.g. for $a=1,2,3,4$, $m_a(29.0)=4$ and $m_a(30.0)=1$ in the subfigure (b). 
}
\label{fig:traj}
\end{figure}
%%%%%

\subsection{Numerical example}
\label{Subsec:NumericalExample}
To show the operation of the population dynamics method,  we introduce a simple model of diffusion %with additive noise 
in a quartic potential.
That is, $F(x)=-x ^3$ and $B(x) = \sqrt{2\epsilon}$, where $\epsilon$ is the noise strength (or temperature).
We take $\lambda_{\rm c}=0$ and $\lambda_{\rm d}=x(x+1)$.  For $h<0$ the distribution $P_h$ is concentrated on trajectories with small values of $\lambda_{\rm d}$,
which tend to
localise near $x\approx-\frac12$.   
Here we focus on the case $h>0$, which leads to unusually large values of $\lambda_{\rm d}$. Those can be realised either for $x>0$ or
$x<0$ but at large $\tau$ this rare event is almost always realised by trajectories that have $x>0$ (as illustrated in Fig.~\ref{fig:traj}). %, as in Fig.~\ref{fig:traj}.
This simple problem can be solved exactly in the zero-noise limit (see Appendix~\ref{appendix:InstantonResults}).

The operation of the population dynamics method is illustrated in Fig.~\ref{fig:traj}.
Fig.~\ref{fig:traj}(a) shows four copies of the system that evolve in time, except that
some trajectory segments stop and others branch, as the cloning operates.  Fig.~\ref{fig:traj}(b) shows four
representative trajectories (sample paths) for the distribution $P_h[X]$, which have been reconstructed from panel (a),
by tracing backwards in time from the clones that survived up to the final time~$\tau$.

\section{Sampling errors within population dynamics}
\label{Sec:SamplingErrors}

\subsection{Distributions $p_{\rm end}$ and $p_{\rm ave}$}
\label{Subsec:pend_pave}

\begin{figure*}
\centering
%\subfigure[~$p_{\rm end}(x)$ for $h=-1$]{
\includegraphics[width=0.67\columnwidth]{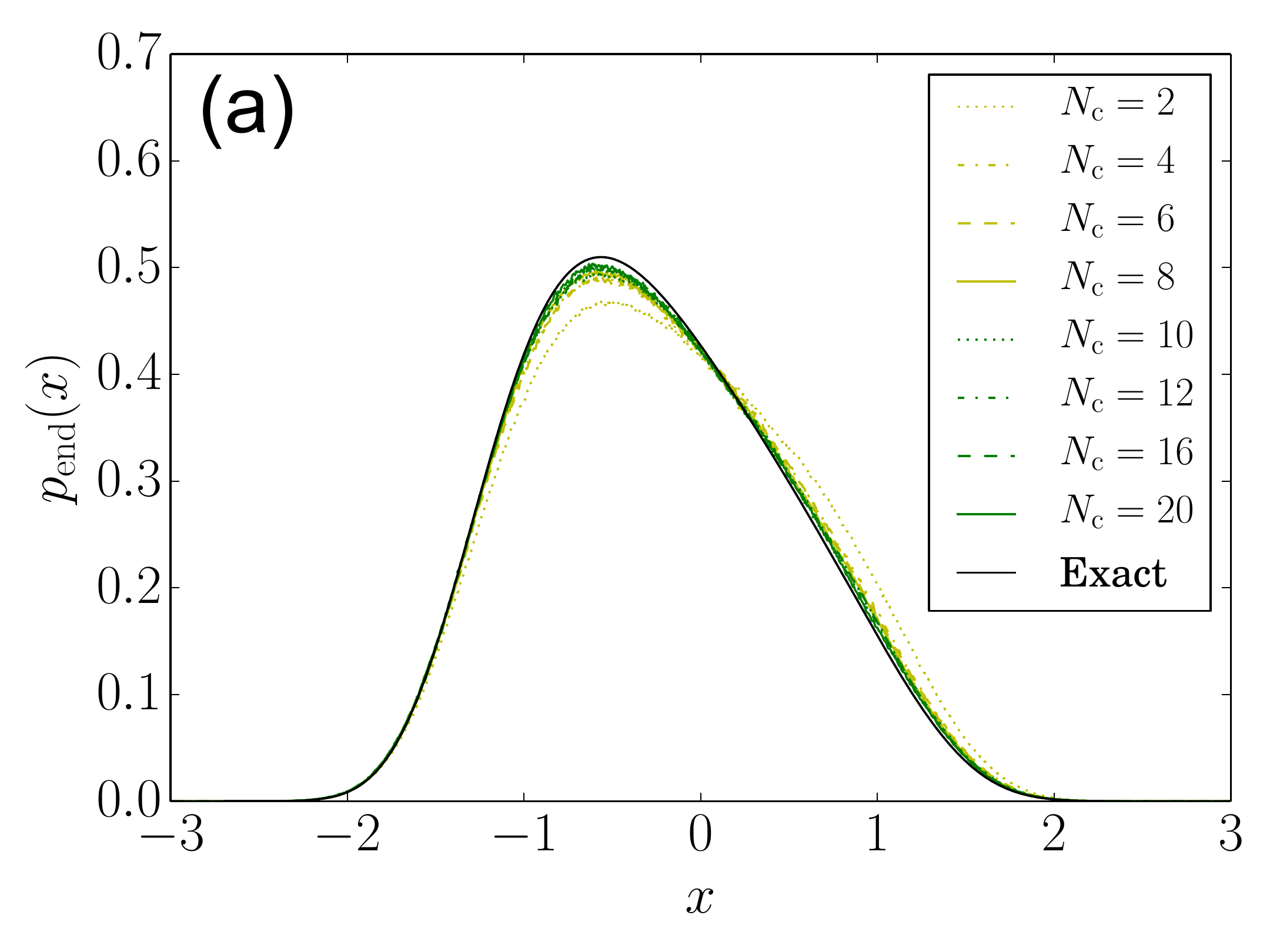}%}
%\subfigure[~$p_{\rm ave}(x)$ for $h=-1$ ($ m_2 = 0.068$, $D_2=0.039$)]{
\includegraphics[width=0.67\columnwidth]{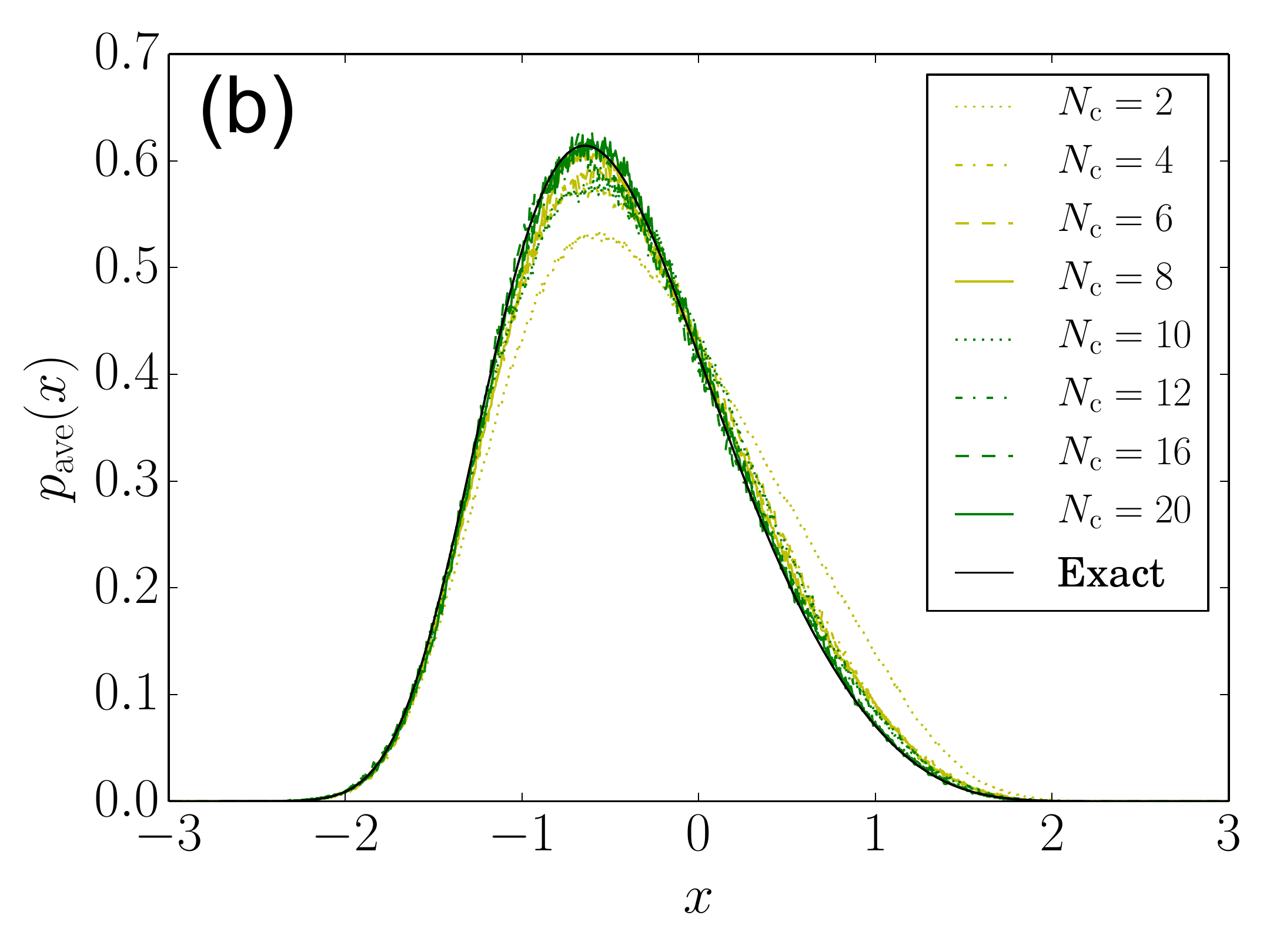}%}
\\
%\subfigure[~$p_{\rm end}(x)$ for $h=1$]{
\includegraphics[width=0.67\columnwidth]{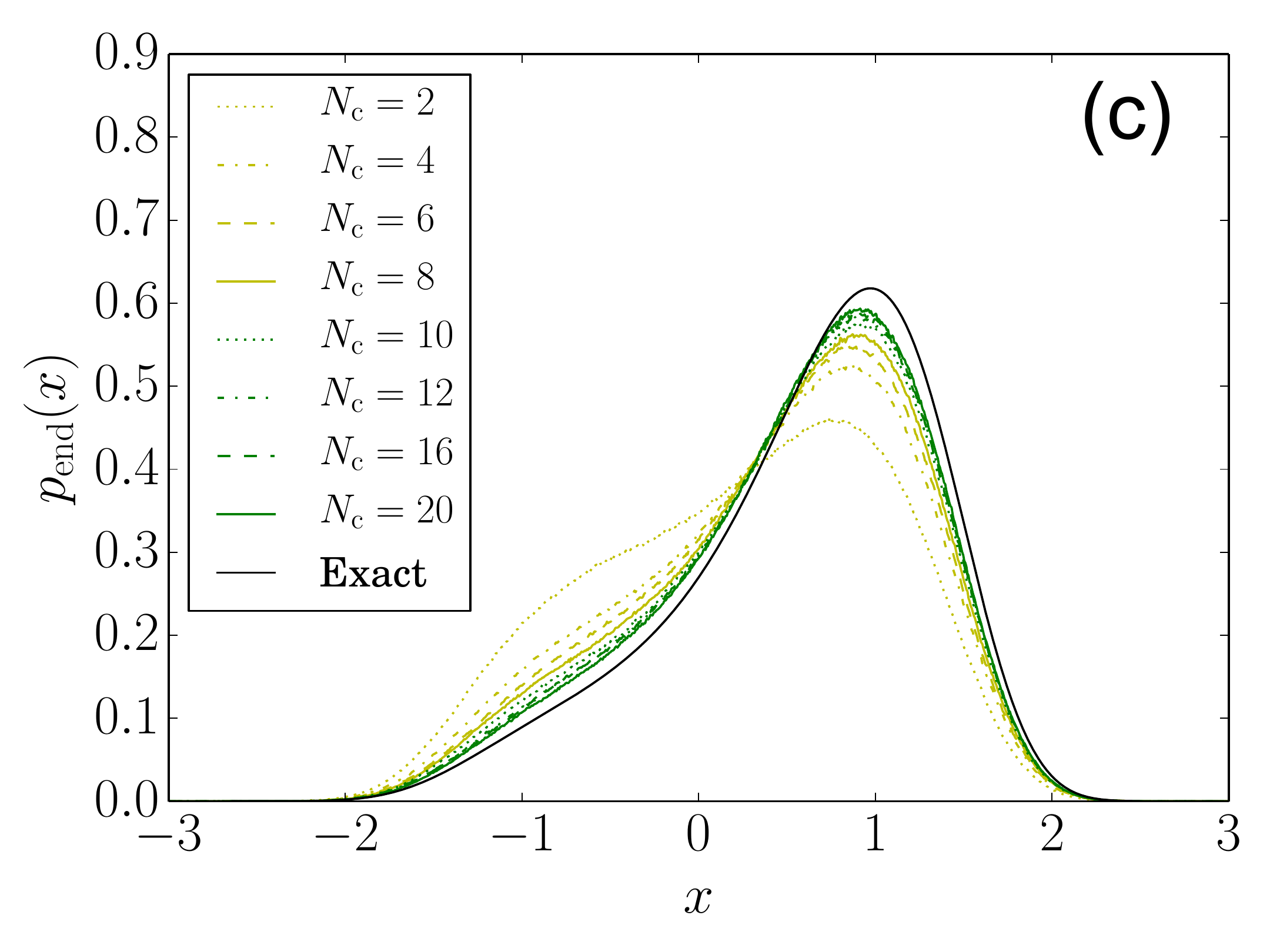}%}
%\subfigure[~$p_{\rm ave}(x)$ for $h=1$ ($m_2 = 0.33$, $D_2=0.17$)]{
\includegraphics[width=0.67\columnwidth]{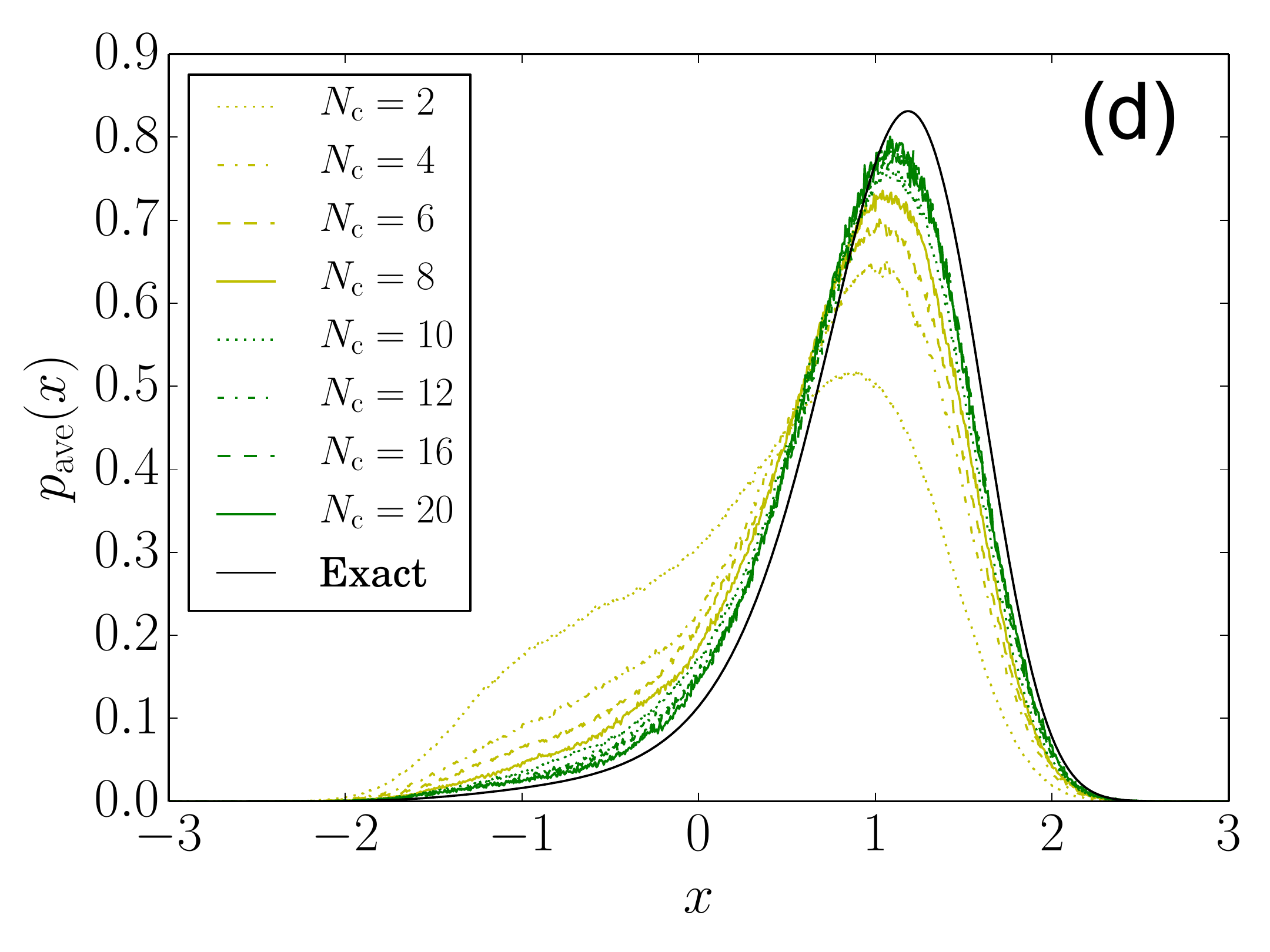}%}
%\subfigure[~$p_{\rm ave}(x)$ for $h=1$, estimation improved by the feedback]{
\includegraphics[width=0.67\columnwidth]{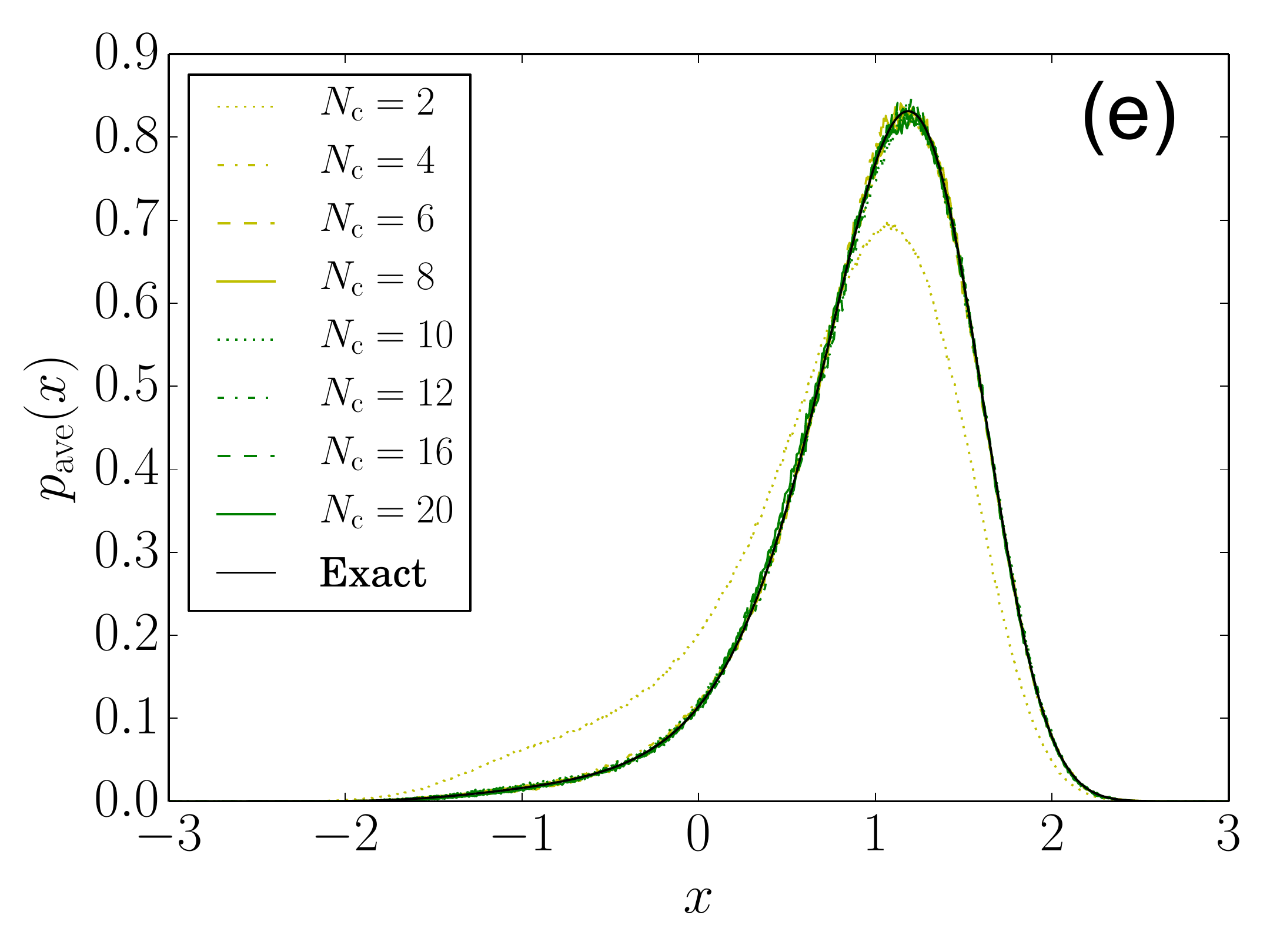}%}
\caption{\label{kappa01ComputationalEfficiencyFeedback} (Color online) \textbf{(a-d)} Distributions $p_{\rm end}(x)$ and $p_{\rm ave}(x)$, defined in~\eqref{avep} and~\eqref{endp}, calculated from the population dynamics method, with various numbers of clones $N_c$. The different panels correspond to a different value of $h$ ($h=\pm 1$) or a different distribution function ($p_{\rm end}(x)$ or $p_{\rm ave}(x)$): (a) $p_{\rm end}(x)$ for $h=-1$, (b) $p_{\rm ave}(x)$ for $h=-1$ , (c) $p_{\rm end}(x)$ for $h=1$, and (d) $p_{\rm ave}(x)$ for $h=1$. For all panels, we set $\epsilon = 1$ and $\tau = 30$.
The numerically exact result is plotted as a black line.   We repeat the simulation $1200/N_c$ times and the result is the average of them (this procedure means that we vary $N_c$ while keeping a fixed computational cost). The results of the population dynamics converge to the analytical ones as $N_c$ increases.
\textbf{(e)} $p_{\rm ave}(x)$ for $h=1$ (improved estimation) calculated from a population dynamics method with control-with-feedback, as described in Section~\ref{Subsec:IteAFeed} and Section~\ref{Sec:NumEx}. 
Results are shown after two iterations of the feedback procedure. 
The exact distribution $p_{\rm ave}(x)$ is again shown as a black line.  The comparison between (d) and (e) indicates
that the convergence with respect to $N_c$ is improved significantly by the control-with-feedback method.
The variance $m_2$  and the relative entropy $D_2$ defined in ~\eqref{equ:m2def} and \eqref{equ:d2def} both measure how much large values of $N_c$ are required for the cloning procedure to be reliable. For the panel (b), (d) and (e), these values are ($ m_2 = 0.068$, $D_2=0.039$), ($m_2 = 0.33$, $D_2=0.17$) and ($m_2 = 0.0064$, $D_2=0.0032$) respectively. 
}
\end{figure*}

%The accuracy of the population dynamics method is limited clones with high {\it multiplicity}, as
%we now explain.
The accuracy of the population dynamics is limited by the number of clones $N_c$ used in its numerical implementation, as we now explain.
Consider the distribution 
\begin{equation}
p_{\rm ave}(x) = \lim_{\tau \rightarrow \infty } \Big\langle  \tau^{-1}\! \int _{0}^{\tau} \delta(x_t - x) \dd t   \Big\rangle_h,
%\frac{\left \langle \left [\frac{1}{\tau} \int _{0}^{\tau} ds\, \delta(x(s) - x) \right ]   e^{ \tau h \Lambda (\tau)}  \right \rangle}{\left \langle  e^{ \tau h \Lambda (\tau)} \right \rangle}.
\label{avep}
\end{equation}
which indicates the fraction of time spent %that the particle spends 
at position $x$, within the biased ensemble.  We also define
\begin{equation}
p_{\rm end}(x) = \lim_{\tau \rightarrow \infty } \left\langle \delta(x_\tau-x)  \right\rangle_h,
%\frac{\left \langle \left [\frac{1}{\tau} \int _{0}^{\tau} ds\, \delta(x(s) - x) \right ]   e^{ \tau h \Lambda (\tau)}  \right \rangle}{\left \langle  e^{ \tau h \Lambda (\tau)} \right \rangle}.
\label{endp}
\end{equation}
which indicates the fraction of trajectories for which the particle's final position is $x$.  For the stationary state of the dynamics~\eqref{equ:dx}, which corresponds to $h=0$, time-translational
invariance ensures that $p_{\rm ave} = p_{\rm end}$.  However, this is not the case for biased ensembles where $h\neq0$, as illustrated in~\cite{Populationdynamics,garrahan2} and in Fig.~\ref{kappa01ComputationalEfficiencyFeedback}.

The population dynamics method provides estimates of both $p_{\rm ave}$ and $p_{\rm end}$. 
Let the position of clone $a$ at time~$t$ be $x^a(t)$, with $a=1\dots N_c$.  Recalling Fig.~\ref{fig:traj}(a), note that the functions $x^a(t)$ are not continuous in time and do not represent sample paths for the distribution $P_h[X]$.
However, from the definition of the population dynamics algorithm (as explained in Appendix~\ref{PDSection_Algorithm}), the distribution of $x^a(t)$ can be used to estimate $p_{\rm end}(x)$, as
\begin{equation}
p_{\rm end}(x) \simeq \frac{1}{\tau N_c} \int_{0}^{\tau}  \sum_{a=1}^{N_c}\delta (x-x^a(t)) \dd t.
\label{eq:pendestimation}
\end{equation} 

In order to construct sample paths, which we denote by $\xt^a(t)$, we trace backwards in time from
the clones that survive up to $\tau$, as shown in Fig.~\ref{fig:traj}(b).  There are still $N_c$ paths $\xt^a$, but these overlap, particularly at early times.  
Since these trajectories correspond to $P_h[X]$, the distribution of $\xt^a$ gives an estimate of $p_{\rm ave}(x)$, as:
\begin{equation}
p_{\rm ave}(x) \simeq  \frac{1}{\tau N_c}  \int_{0}^{\tau}   \sum_{a=1}^{N_c} \delta (x-\xt^a(t)) \dd t.
\label{eq:paveestimation}
\end{equation}
 The approximate equalities in the relations (\ref{eq:pendestimation}) and (\ref{eq:paveestimation}) become exact in the limit $N_c\to\infty$ and $\tau \rightarrow \infty$, in which the population dynamics gives exact results.

We show numerical examples of these functions in Fig.~\ref{kappa01ComputationalEfficiencyFeedback},
for a particle moving in a quartic potential, as introduced in Section~\ref{Subsec:NumericalExample}. 
%(That is, $d=1$, $F(x) = - x^3$, $B(x)=\sqrt{\epsilon}$, $\lambda_{\rm d}(x) = \lambda(x) \equiv x^2 + x$, and $\lambda_{\rm c}(x) = 0$.)
We estimate $p_{\rm ave}$ and $p_{\rm end}$ from (\ref{eq:pendestimation}) and (\ref{eq:paveestimation}), and show them in Fig.~\ref{kappa01ComputationalEfficiencyFeedback}. In the same figure, we also plot the numerically exact distributions, obtained by numerical solution of a modified Fokker-Planck equation (see~\cite{Touchette} and Appendix~\ref{sec:fp}).  The population dynamics converges to the exact result as $N_c$ is increased.  Also shown in Fig.~\ref{kappa01ComputationalEfficiencyFeedback} are results using the control-with-feedback method that we introduce in this paper: these results will be discussed in later sections.

\subsection{Multiplicity}
\label{subsec:Multiplicity}

The population dynamics method gives accurate results in the limit of large $N_c$.  The central idea is that in a large population, short-lived rare fluctuations will occur.  Based on these short-lived fluctuations, we duplicate some of the clones: repeated application of this procedure generates the \emph{long-lived} fluctuations that are relevant for large deviation theory.  For this to be effective, the population on which the cloning operates must be large enough to capture the relevant short-lived fluctuations.  That is, the cloning part of the algorithm can allocate extra statistical weight to configurations that are already present in the population, but new configurations are only generated by the natural (unbiased) dynamics of the system.

Assuming that $N_c$ is large enough for efficient operation of the algorithm, the configurations that are associated with long-lived dynamical fluctuations are distributed as $p_{\rm ave}$, but the cloning step operates on a population distributed as $p_{\rm end}$.  From the argument above, it is clear that if \emph{typical} samples from $p_{\rm ave}$ are rare with respect to $p_{\rm end}$, then a large population is required in order to obtain accurate results.  To quantify this, it is useful to define
the multiplicity $m_a(t,\tau)$ of clone $a$ at time $t$ as the number of its descendants that survive until the final time $\tau$ (see Fig.~\ref{fig:traj}). 
Rewriting (\ref{avep}) as $p_{\rm ave}(x) \simeq  \frac{1}{\tau N_c} \int_{0}^{\tau} \sum_{a=1}^{N_c} m_a(t,\tau) \delta (x-x^a(t)) \dd t$ and comparing with (\ref{endp}),
one sees that for a clone with position $x=x^a(t)$, the \emph{expected value} of its future multiplicity is $p_{\rm ave}(x)/p_{\rm end}(x)$.  Since the clone positions $x^a(t)$ are distributed as $p_{\rm end}$, averaging this future multiplicity over configurations $x$ yields $\int p_{\rm end}\cdot(p_{\rm ave}/p_{\rm end})\mathrm{d}x=\int p_{\rm ave}(x)\mathrm{d}x =1$, which reflects the fact that the population size is constant in time.
% Comparing (\ref{avep},\ref{endp}), the average multiplicity of a clone at position $x$ is proportional to $p_{\rm ave}(x)/p_{\rm end}(x)$.

\begin{figure}
\centering
%\subfigure[~$N_c^\prime(t)$]{
%\includegraphics[width=\columnwidth]{comparisons-controlled-non-controlled}%}
%\includegraphics[width=0.95\columnwidth]{G_hfeedbackpopdynZoomup}\\%}
%\subfigure[~$\tilde N_c(t)$ for $h=-1$]{
\includegraphics[width=0.9\columnwidth]{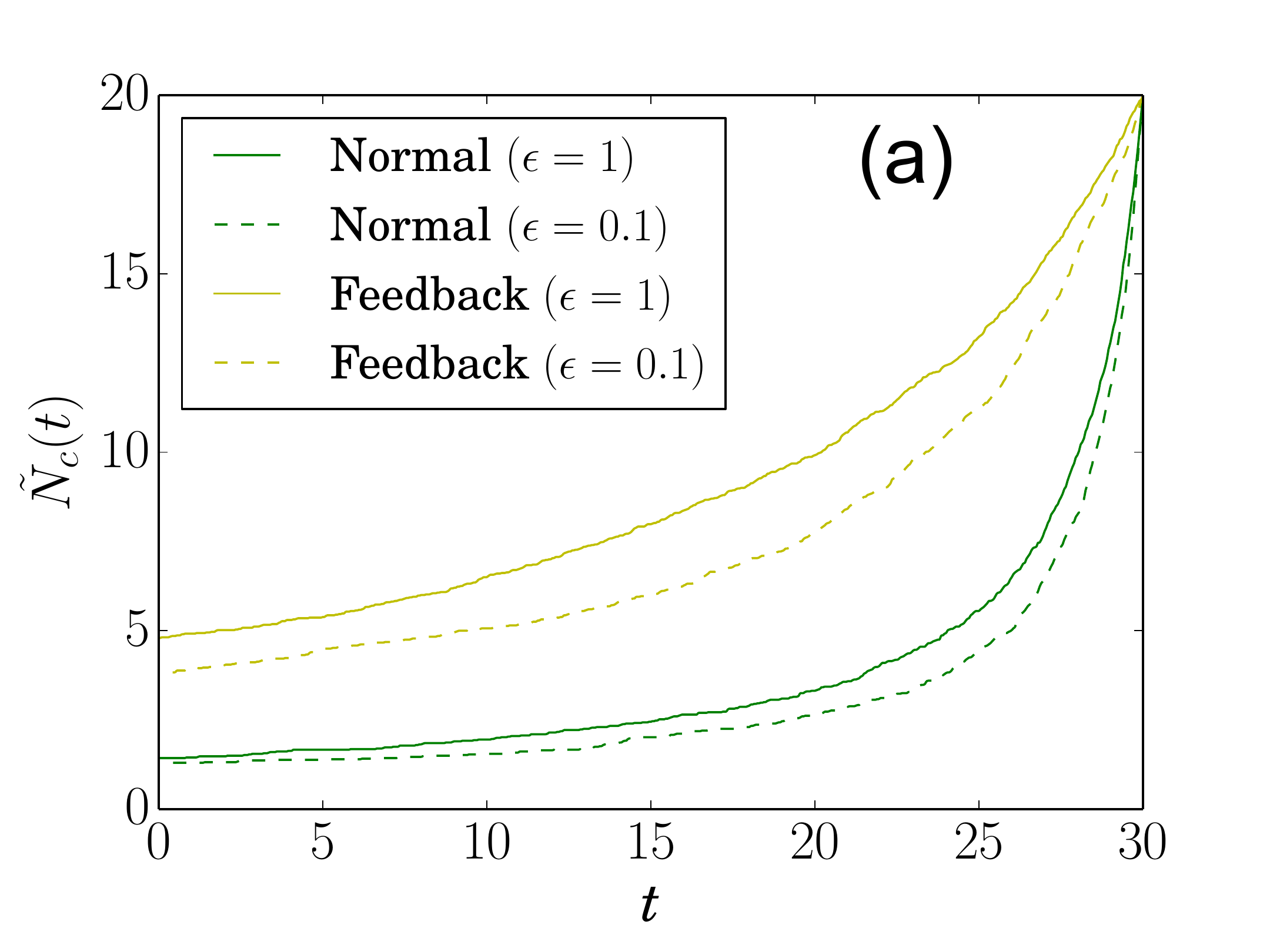}%}
\\
%\subfigure[~$\tilde N_c(t)$ for $h=1$ ]{
\includegraphics[width=0.9\columnwidth]{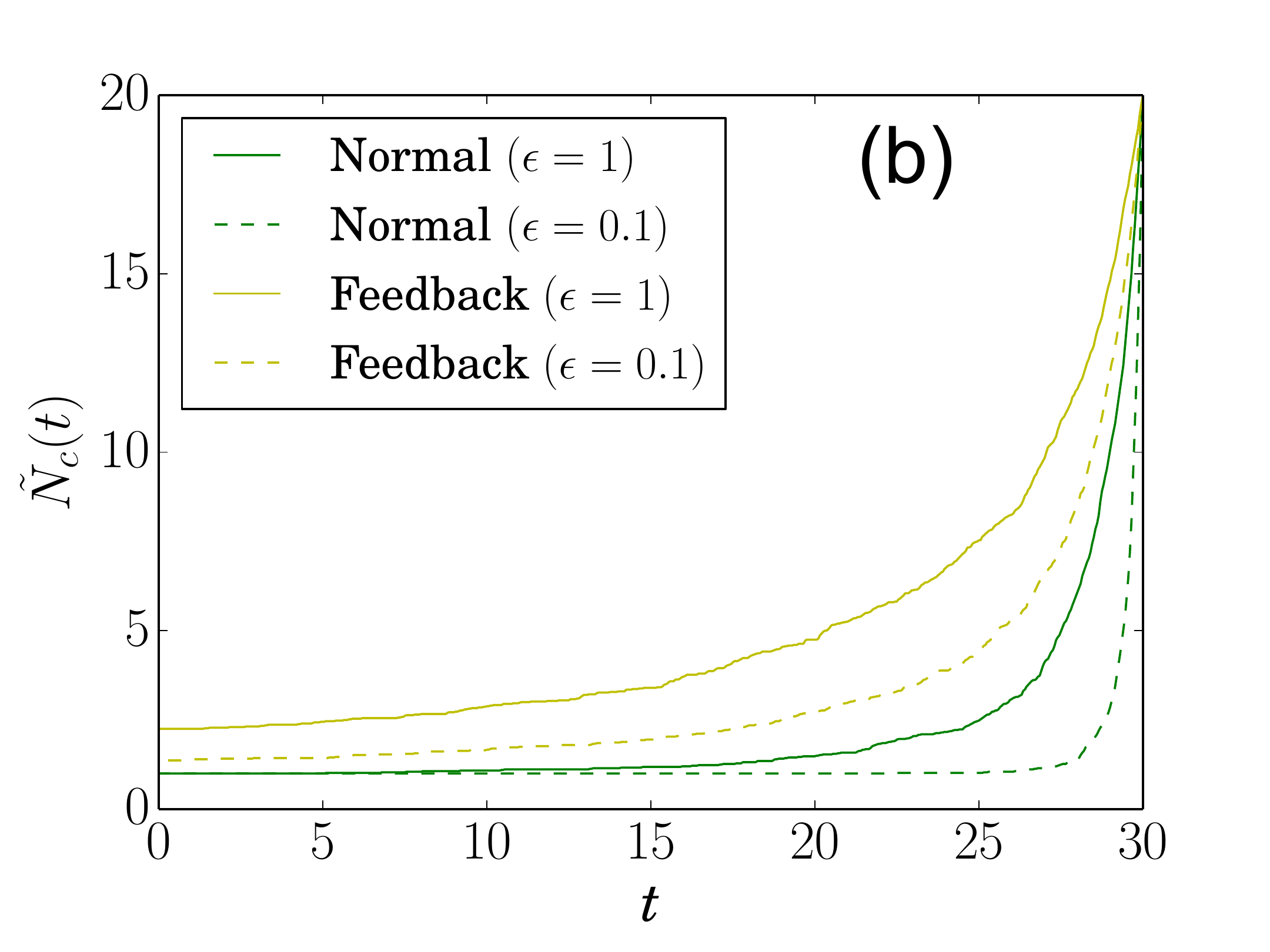}%}
\caption{%\label{dd1_01hhm1Numofcopies_NormalAndAuxi} 
(Color online) The number of independent (distinct) clones $\tilde N_c(t)$ obtained from the normal population dynamics method (green line or dark grey line in the printed version) for $h=-1$ (a) and $h=1$ (b). 
The line type corresponds to the value of noise intensity: $\epsilon = 1$ (solid line) and $\epsilon = 0.1$ (dashed line). 
We set $N_c=20$ and $\tau=30$. 
When the distributions $p_{\rm ave}$ and $p_{\rm end}$ are very different from each other, we expect that $\tilde N_c(t)$ decreases rapidly as $t$ decreases from $\tau$: to illustrate this, note that (for $\epsilon =1$)  $m_2 = 0.068$ for $h=-1$ and $m_2 = 0.33$ for $h=1$: the same ordering is preserved for smaller $\epsilon$.
We also plot  $\tilde N_c(t)$ obtained from the controlled population dynamics (yellow line or light grey line in the printed version) with the control-with-feedback explained in Section~\ref{Subsec:IteAFeed} and Section~\ref{Sec:NumEx}. 
The larger values of $\tilde N_c(t)$ obtained with the control-with-feedback method lead to smaller statistical uncertainties in the results.
}
\label{fig:NtGh}
\end{figure}

In practice, the distribution of the multiplicity $m_a(t,\tau)$ is very broad, and \emph{typical} multiplicities are far from their \emph{average} values.  There are many clones for which no descendants survive until time $\tau$ (see Fig.~\ref{fig:traj}(a)), in which case $m_a(t,\tau)=0$.  In order to maintain an average multiplicity of $1$, these zero-multiplicity clones are balanced by a small number of clones with larger multiplicity.  It is useful to define $\tilde{N}_c(t,\tau)$ as the number of clones that are present in the population at time $t$, for which $m_a(t,\tau)>0$.  
Numerical results for $\tilde N_c(t,\tau)$ are shown in Fig.~\ref{fig:NtGh} -- this quantity decreases rapidly as $t$ decreases away from $\tau$, showing that many clones have no surviving descendants: it follows that the multiplicities of the surviving clones must be large.
From (\ref{eq:paveestimation}), one sees that if $\tilde{N}_c(t,\tau)$ is small, numerical estimates of $p_{\rm ave}$ contain only a small number of independent samples, which can lead to large numerical uncertainties within the algorithm.

Moreover, the presence of large multiplicities within the cloning scheme can lead to large \emph{systematic errors}, which cannot be reduced by averaging over repeated runs of the same algorithm.  On running the system with a fixed population, the future multiplicity of any clone is bounded above by the population size $N_c$.  We will show in the next section that this constraint has serious implications for systems in the small noise limit.  More generally, in order to characterise whether a system requires a large population or not, it is useful to define two numbers that measure how different are the distributions $p_{\rm ave}$ and $p_{\rm end}$.  These are
\begin{equation}
m_2 = \int  p_{\rm end}(x) \left[\left(\frac{p_{\rm ave}(x)}{p_{\rm end}(x)}\right)^2 - 1\right] \dd x
\label{equ:m2def}
\end{equation}
and
\begin{equation}
D_2 = \int p_{\rm ave}(x) \log \left(\frac{p_{\rm ave}(x)}{p_{\rm end}(x)}\right)\dd x.
\label{equ:d2def}
\end{equation}
Given that $p_{\rm ave}(x)/p_{\rm end}(x)$ is the expected future multiplicity of a clone at $x$, we recognise $m_2$ as the variance of this multiplicity, with respect to the distribution $p_{\rm end}$ of clone positions (recall that the average multiplicity with respect to this distribution is equal to unity).  Similarly $D_2$ is the relative entropy of $p_{\rm ave}$ with respect to $p_{\rm end}$~\cite{footnote:KullbackLeibler}: this is related to the controlling forces that will be introduced in Section~\ref{Sec:FB_PD}. 
Large values of $m_2$ and $D_2$ indicate that $p_{\rm end}$ and $p_{\rm ave}$ are different from each other, in which case larger values of $N_c$ will be required for accurate results within population dynamics.  For the two cases $h=\pm1$ shown in Fig.~\ref{kappa01ComputationalEfficiencyFeedback}, we have for $h=-1$ that $(m_2,D_2)=(0.068,0.039)$ while for $h=+1$, $(m_2,D_2)=(0.33,0.17)$, reflecting the larger populations required for accurate results when $h=+1$.  Obtaining general estimates of the actual population size $N_c$ required for convergence is an important goal for future work.

\subsection{Sampling problems for weak noise}
\label{subsec:SmallNoiseProblem}

The effect described in the previous section is particularly severe for systems where the random (noise) force in (\ref{equ:dx}) is small.  To illustrate this case, we set $B(x) = \sqrt{2\epsilon} B_0(x)$, consistent with the numerical example of Sec.~\ref{Subsec:NumericalExample} (for which $B_0=1$).  The small noise limit is then $\epsilon\to0$.
%When the noise intensity is small, the population dynamics suffers a serious convergence problem with respect to $N_c$.
We define $x^*=\argmax_x [p_{\rm ave}(x)]$ as the most likely value of $x$, within the distribution $p_{\rm ave}$.  The population dynamics requires that the typical multiplicity of a clone with position $x^*$ should be (at least) of order 
$m^*\equiv p_{\rm ave}(x^*)/p_{\rm end}(x^*)$. This clearly cannot be achieved unless $N_c \gtrsim m^*$, which provides an estimate of the number of clones required for accurate results. 

This multiplicity $m^{*}$ increases exponentially as the noise intensity of the system becomes small. 
In this limit, the dynamics of the 
system runs increasingly slowly so it is natural to rescale either the time variable or (equivalently) the biasing field $h$ as
$\tilde G(\tilde h) \equiv \epsilon G(h)$ with $\tilde h \equiv h \epsilon$. (This scaling also appears in the hydrodynamic limit of microscopic models~\cite{Kipnis_Landim}.)
In this limit, $p_{\rm ave}$ and $p_{\rm end}$ satisfy a large deviation principle with respect to the noise intensity $\epsilon$: $p_{\rm ave}(x) \sim \ee^{- I_{\rm ave}(x)/\epsilon} $ and $p_{\rm end}(x) \sim \ee ^{- I_{\rm end}(x)/\epsilon}$. Hence,
$m^* \sim \ee ^{  I_{\rm end}(x^{*}) /  \epsilon}$,
where we used $I_{\rm ave}(x^{*}) = 0$. This indicates that we need an exponentially large $N_c$ as $\epsilon$ becomes small. More quantitatively, we define a characteristic noise intensity $\epsilon^*$ by
\begin{equation}
\epsilon^* \equiv \frac{1}{  I_{\rm end}(x^{*}) }.
\label{eq:defepsilonstar}
\end{equation}
For $\epsilon<\epsilon^*$, we expect that population dynamics can not be used practically, because of the exponentially large $N_c$ required.

\begin{figure}
\centering
%\subfigure[~$N_c^\prime(t)$]{
%\includegraphics[width=\columnwidth]{comparisons-controlled-non-controlled}%}
\includegraphics[width=0.95\columnwidth]{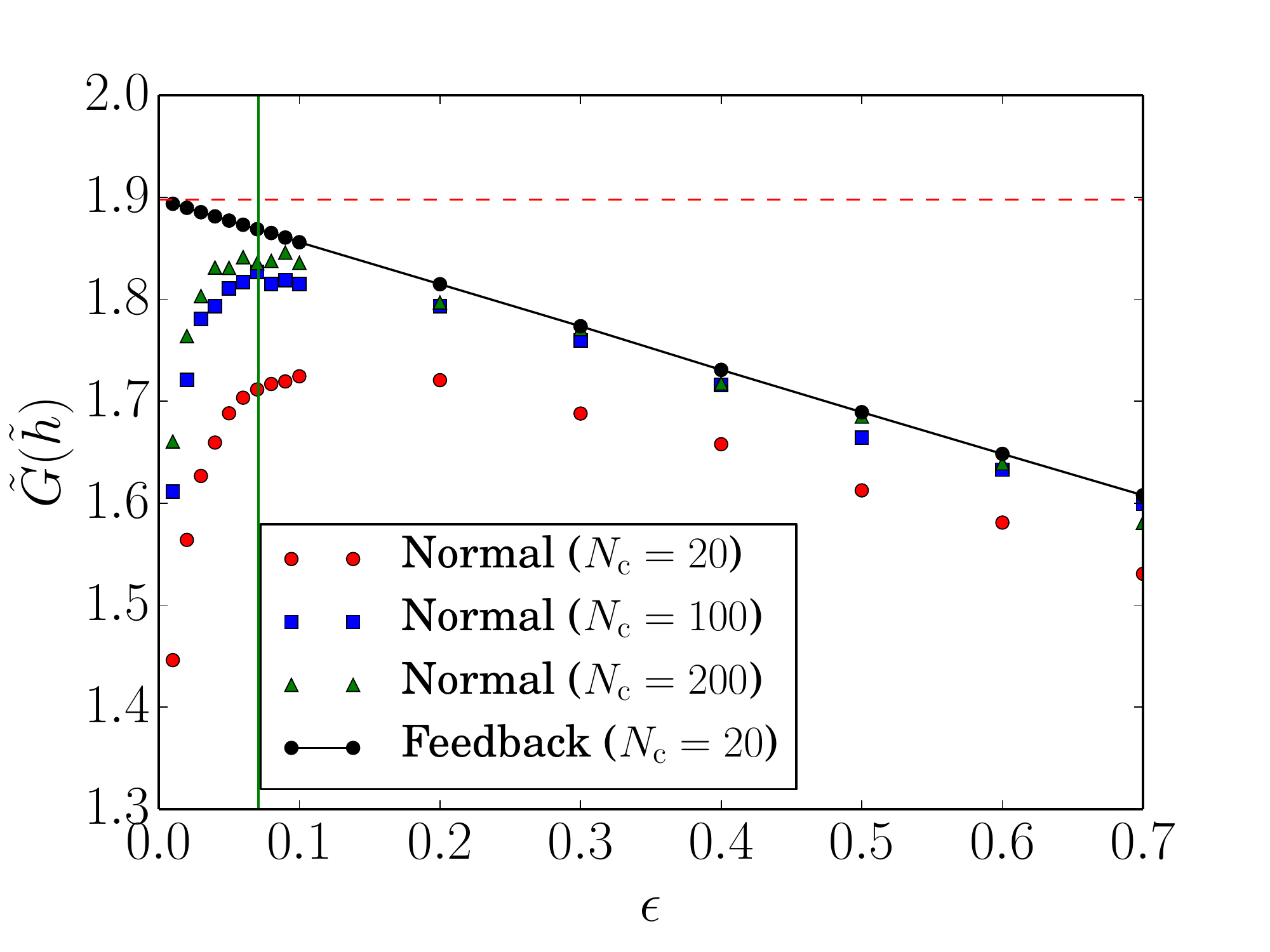}\\%}
%\subfigure[~$\tilde G(\tilde h)$ as a function of $\epsilon$]{
\caption{%\label{dd1_01hhm1Numofcopies_NormalAndAuxi} 
(Color online) Estimates of $\tilde G(\tilde h=1)$, as $\epsilon$ is varied.   We compare results from
the normal population dynamics and from the control-with-feedback method explained in Section~\ref{Subsec:IteAFeed} and Section~\ref{Sec:NumEx}. The analytical result for $\lim_{\epsilon \rightarrow 0} \tilde G(\tilde h)$  is shown as a red dashed line, and the characteristic value of the noise intensity $\epsilon ^*$, defined in~\eqref{eq:defepsilonstar}, is plotted as a green vertical solid line. 
The standard method fails for $\epsilon$ smaller than $\epsilon^*$ but the control-with-feedback method (black continuous line and black circles) converges to the correct value even for $\epsilon < \epsilon ^*$.
}
%\label{fig:NtGh}
\label{fig:Gh_eps}
\end{figure}

As a numerical example, we again consider
the Brownian particle introduced in Section~\ref{Subsec:NumericalExample}.
We numerically estimate $\epsilon^*$ by using a quadratic approximation of the large deviation function $I_{\rm end}(x)$. We plot it as a green vertical line in Fig.~\ref{fig:Gh_eps}. In the same figure, we show the result of the population dynamics for $\tilde G(\tilde h)$ as $\epsilon$ is reduced, with a red constant line corresponding to the analytical value of $\tilde G(\tilde h)$ in the $\epsilon \rightarrow 0$ limit (See Appendix~\ref{appendix:subsectionCumulant} for its determination). Below the characteristic value~$\epsilon^*$, the population dynamics method converges very poorly  as $N_c$ increases.

\section{Population dynamics with a feedback control}
\label{Sec:FB_PD}

\subsection{Controlled dynamics} 
\label{subsec:ControlledDynamics}
To resolve the sampling issues described in the previous section, we introduce a ``control strategy'', which modifies the original
model (\ref{equ:dx}), in order to make the rare events of interest more likely. (These large deviation problems have dual representations in terms of optimal control problems~\cite{OptimalControlFleming, jack15-epj,chetrite15-jstat,fleming92,hartmann12,kappen13}, which provide a natural interpretation of the method presented here.)
The modified model is
\begin{equation}
\dot{x}_t = F(x_t) + w(x_t) + B(x_t) \xi_t,
\label{modifiedsystemdef}
\end{equation}
where $w(x)$ is a controlling force which we write as 
\begin{equation}
w(x) = h \kappa \lambda_c(x)    - \kappa \nabla V(x), 
\end{equation}
where $V$ acts as a potential.
A straightforward calculation shows that
averages with respect to the biased distribution $P_h$ can be rewritten as averages with respect to this modified dynamics, but with a bias on a different observable $\Lambda^w$,
which replaces $\Lambda$.  That is, by defining 
\begin{equation}
\Lambda^w = \frac{1}{\tau}\int_0^\tau \lambda^w(x_t) \dd t
\end{equation}
 with
\begin{equation}
\lambda^w = \lambda_{\rm d}  +\frac 1h[ (F+w/2)\cdot \kappa^{-1} w - \tfrac{1}{2} \mathrm{Tr}( H_V \kappa) ],
\end{equation}
% Note the initial condition $\pi_0$ used in $P_w$ is the same
% as that used in $P_h$.
%, which is the steady state distribution for the original process (\ref{equ:dx}).
in which $H_V$ is a Hessian matrix with elements $(\partial^2 V/\partial x_i \partial x_j)$,
we have
% (\emph{RLJ: I hope this is all true!})
\begin{equation}
P_h[X] = P_w[X] \ee^{ V(x_\tau) - V(x_0)} 
\label{equ:phw}
\end{equation}
with 
\begin{equation}
P_w[X] = \pi_0(x_0) \exp\!\Big[\!-\!\!\int_0^\tau\!\!\! {\cal L}^w(x_t,\dot{x}_t)  \dd t + h \tau \Lambda^w(\tau) \Big],
\end{equation}
%\begin{align}
%P_h[X] &= P_w[X] \ee^{ V(x_\tau) - V(x_0)} 
%\qquad \text{with}
%\label{equ:phw}
%\\
%P_w[X] &= \pi_0(x_0) \exp\!\Big[\!-\!\!\int_0^\tau\!\!\! {\cal L}^w(x_t,\dot{x}_t)  \dd t + h \tau \Lambda^w(\tau) \Big]
%\label{equ:pw}
%\end{align}
where ${\cal L}^w$ is the action corresponding to the controlled Langevin equation (\ref{modifiedsystemdef}) obtained by replacing $F\mapsto F+w$ in~\eqref{eq:Lagrangian}.  See Appendix~\ref{Sec:App_Derivation} for details of the derivation. We stress that these relations are satisfied for {\it any} control $w$.

Averages with respect to $P_w$ are denoted by $\langle \cdot \rangle_w$,
and can be calculated using the population dynamics method with the modified model~(\ref{modifiedsystemdef}).  
Physically, equation (\ref{equ:phw}) says that rare events for the system (\ref{equ:dx}) have an alternative characterisation as 
rare events for the controlled system (\ref{modifiedsystemdef}). More precisely,   
from (\ref{equ:phw}), the averages $\langle \cdot \rangle_h$ and $\langle \cdot \rangle_w$
are not equal, but their associated probability distributions differ only through boundary terms at initial and final times. 
For large $\tau$, we focus on properties far from initial and final times, in which case
the averages  $\langle \cdot \rangle_h$ and $\langle \cdot \rangle_w$ are equivalent. 
This equivalence implies that 
\begin{equation}
p_{\rm ave}^{w} = p_{\rm ave},
\label{eq:pave_pavew}
\end{equation}
where $p_{\rm ave}^{w}$ is defined as in (\ref{avep}) but for the controlled population dynamics~\eqref{modifiedsystemdef}.  
On the other hand, when we consider properties close to the final time $\tau$ (which are not relevant for the large deviations 
of time-averaged quantities), the two averages $\langle \cdot \rangle_h$ and $\langle \cdot \rangle_w$ are different in general. For example,
the end-time distribution $p_{\text{end}}^w$ for the controlled dynamics differs from its uncontrolled counterpart as
\begin{equation}
p_{\rm end}^{w} \propto p_{\rm end} \ee^{-V(x)}, 
\label{eq:pend_pendw}
\end{equation}
as read from~\eqref{equ:phw} (or see Appendix~\ref{sec:fp} for a detailed derivation of~\eqref{eq:pave_pavew} and~\eqref{eq:pend_pendw}). 
Thus the control $w$ allows $p_{\rm end}^w$ to be varied, while always keeping $p_{\rm ave}^w$ constant (and hence leaving unchanged the bulk properties of $P_h$, which are relevant for the large deviations of time-averaged quantities).

\subsection{Optimal control} 
These results apply for any control force $w$, but a (unique)
 optimal choice $w^*$ can be defined by the condition 
\begin{equation}
p_{\rm ave}^{w^*}=p_{\rm end}^{w^*}. 
\label{eq:def_optimalcontrol}
\end{equation}%(meaning that the typical $m_a$ is 1 for any $a$). 
From (\ref{equ:m2def},\ref{equ:d2def}), this result implies that for the controlled population dynamics, $m_2=D_2=0$: all clones have expected future multiplicity of unity, regardless of their position.  In fact, this case also implies that $\lambda^{w^*}(x)$ is independent of $x$ (see  Appendix~\ref{sec:fp}), 
so that there is no cloning or deletion of clones in the optimally-controlled population dynamics algorithm. That is, all multiplicities are equal to unity (not just their expected values).  
The result is that the optimally-controlled process~\cite{jack15-epj,chetrite15-jstat,fleming92,hartmann12,kappen13} generates directly the path measure $P_h$, up to the corrections given in~(\ref{equ:phw})~\cite{Evans,JackSollich,NemotoSasa2,maes08,Chetrite_Touchette2}.  
Note also that $D_2$, as defined in (\ref{equ:d2def}) for the original population dynamics, is also related to an average of the optimal control potential $V^*$ (where $V^*$ is the potential $V$ corresponding to the optimal control $w^*$), since $\log [p_{\rm ave}(x)/p_{\rm end}(x)] = -V^*(x) - \log [\int \ee^{-V^*(x')} p_{\rm end}(x')\mathrm{d}x']$.  
%The force $w^*$ can also be obtained from optimal control theory~\cite{jack15-epj,chetrite15-jstat}.  

The optimal control can be estimated by using the population dynamics with {\it any} non-optimal control force $w$ (or its corresponding potential $V$).
We perform the population dynamics and generate sample paths from $P_w$.  
From the definition of the optimal force (\ref{eq:def_optimalcontrol}) with the relations between $p_{\rm end, ave}^w$ and $p_{\rm end, ave}$ (\ref{eq:pave_pavew}), (\ref{eq:pend_pendw}), we obtain
\begin{equation}
 V^*(x) =  V(x) + \log \frac{p^w_{\rm end}(x)}{ p^w_{\rm ave}(x)}.
\label{equ:pwv}
\end{equation}
%where $V^*$ is the potential $V$ corresponding to the optimal control $w^*$.   
Since all terms on the right-hand side of (\ref{equ:pwv}) can be measured from the population dynamics with a non-optimal control $w$, this allows an estimate of $V^*$, and hence of~$w^*$.

\subsection{Control-with-feedback for population dynamics}
\label{Subsec:IteAFeed}  %% RLJ moved this label....

Based on (\ref{equ:pwv}), we arrive at the following iteration and feedback scheme for efficient analysis of large deviations of $\Lambda(\tau)$.  Starting with the original population dynamics of~\cite{Populationdynamics}, we obtain estimates $p_{\rm end}^0$ and $p_{\rm ave}^0$ of $p_{\rm end}$ and $p_{\rm ave}$, and we use (\ref{equ:pwv}) to obtain an estimate of the optimal control potential $V^*$, which we denote by $V^1$.  We then repeat the population dynamics calculation with a control force $w=w^1$ derived from the potential $V^1$. We use results from this new calculation together with (\ref{equ:pwv}) to obtain a new (more accurate) estimate of the optimal control.  Iterating this scheme, the estimate of $V^*$ at iteration $r$ is $V^r$.  As $V^r\to V^*$, we have from (\ref{equ:pwv}) that $p^w_{\rm end} \to p^w_{\rm ave}$, and hence the sampling problems described in Sec.~\ref{subsec:Multiplicity} are reduced.  This improves the accuracy of the population dynamics method.

Given sufficiently many clones $N_c$, the original method of~\cite{Populationdynamics} can already provide accurate results, but we have demonstrated that for finite $N_c$ there may be large systematic errors.  The strength of our scheme is that on repeated iteration, the control potential $V$ approaches the optimal control $V^*$, and the errors within the method are reduced.  Thus, the numerical accuracy of the method increases as the scheme is iterated.

\newcommand{\fn}{\zeta}
\newcommand{\cut}{k}

For the implementation of this iteration scheme, we require a computational representation of the function $V(x)$, and its gradient $\nabla V$.  From (\ref{equ:pwv}), a natural choice might be to represent $p_{\rm ave}$ and $p_{\rm end}$ by histograms, based on a discretisation of the configuration space.  However, this choice does not facilitate estimation of $\nabla V$, and it is also unfeasible in high-dimensional systems.  We therefore use a potential $V$ that is defined in terms of a set of basis functions $\fn_i$, with coefficients $c_i$:
\begin{equation}
V(x) = \sum_{i=1}^{\cut} c_i \fn_i(x) .
\label{eq:effective_definition}
\end{equation}
where $\cut$ is the size of the basis set.

%In numerical simulations, it is computationally demanding to divide the full state space %(phase space) 
%into small bins and store the values of $V(x)$, especially in many-body and high-dimensional systems. 
%Instead, we use a representation of $V(x)$ as an expansion, such as
%\begin{equation}
%V(x) \sim \sum_{i=0}^{\theta} c_i^{w} \xi_i(x),
%\label{eq:effective_definition}
%\end{equation}
%where $\xi_i(x)$ is a fixed set of functions to expand $V(x)$, $\theta$ is a truncating number in the expansion, and $(c_i^{w})_{i=0}^{\theta}$ are constants. 

At stage $r$ of our iterative scheme, the coefficients $c$ are denoted by $\bm{c}^r = (c^r_i)_{i=1}^{\cut}$.  In the absence of prior information about the optimal control $V^*$, the first stage of the method ($r=0$) uses the original population dynamics, so $c_i^0=0$ for all $i$. In stage $r+1$, we update these coefficients according to (\ref{equ:pwv}) so that the potential $V^{r+1}$ in the next stage is the best available estimate of $V^*$.  There is considerable freedom in how to obtain this estimate: we take
\begin{multline}
\bm{c}^{r+1} = {\rm argmin}_{\bm{c}} \int _{\Omega_r}  \Bigg[ V^r(x) + \log \frac{p^{w,r}_{\rm end}(x)}{p^{w,r}_{\rm ave}(x)} 
\\ -  \sum_{i=0}^{\cut} c_i \fn_i(x) \Bigg]^2 \dd x,
\label{eq:sim_definition}
\end{multline}
where $p^{w,r}_{\rm end}$ is the numerical estimate for $p^w_{\rm end}$ obtained at iteration $r$, and similarly $p^{w,r}_{\rm ave}$. The state space $\Omega_r $ is defined as the space where $p^{w,r}_{\rm ave}>0$ (note that $p^{w,r}_{\rm end}(x)>0$ whenever $p^{w,r}_{\rm ave}(x)>0$, from the definition of how to construct $\tilde x^a(t)$ as shown in Fig.~\ref{fig:traj}(b)).

%Here, we assume that there is a unique way to construct the coefficients
%$(c_i^{w})_{i=0}^{\theta}$ from $V(x)$, which is symbolised as $\sim$ in (\ref{eq:effective_definition}). There are several ways to define this symbol $\sim$, for example, such as the minimizer of the $L^{2}$ distance between the left-hand side and the right-hand side: 
%\begin{equation}
%(c_i^{w})_{i=0}^{\theta}  \equiv {\rm Argmin}_{(c_i^{w})_{i=0}^{\theta}}\int dx \left [ V(x) -  \sum_{i=0}^{\theta} c_i^{w} \xi_i(x) \right ]^2,
%\label{eq:sim_definition}
%\end{equation}
%which is indeed the definition used in the numerical example in Section~\ref{Sec:NumEx}.

We emphasise that, for {\it any} basis set $\fn_i$ (with {\it any} truncation number $\cut$), eq.~(\ref{equ:phw}) is satisfied, meaning that if the number of clones $N_c$ and the time $\tau$ are large enough, the result of {\it any}  controlled population dynamics always leads to the same results, which can also be obtained from the original (uncontrolled) population dynamics.  However,
the choice of the expansion functions $\fn_i(x)$ (and the value of the truncation number $\cut$) does affect the computational cost, through the number of clones required for convergence, as discussed in Section~\ref{subsec:Multiplicity}.

\subsection{Advantages of the control-with-feedback for population dynamics, and relation to other methods}
\label{Subsec:whyuseful}

\begin{figure}
\centering
\includegraphics[width=1.\columnwidth]{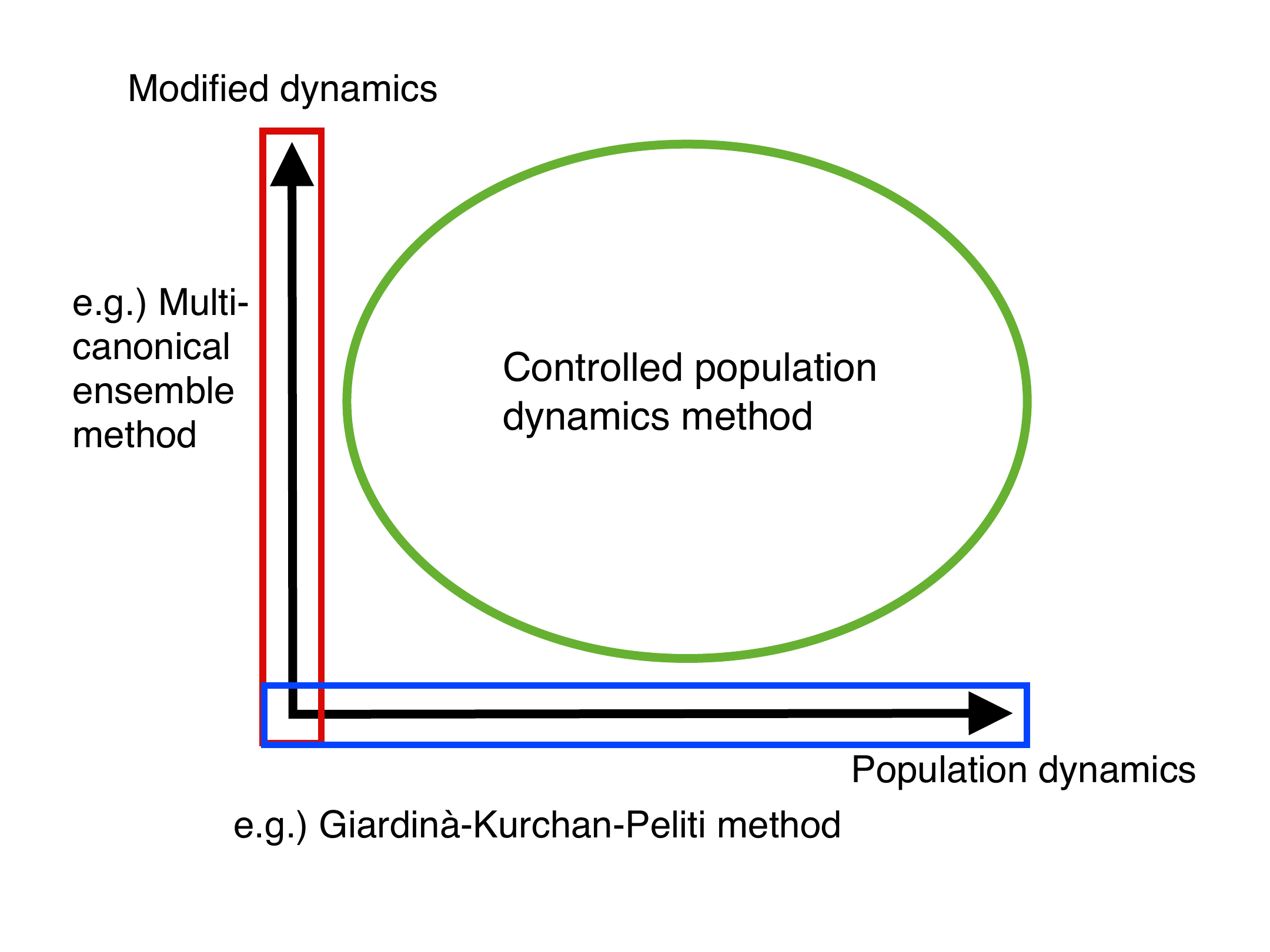}
\caption{
(Color online) A schematic map illustrating the methodological situation of the controlled population dynamics. %By considering the optimal control, the method gets on the $y$-axis, while by setting $V=0$, the method gets on the $x$-axis.
}
\label{fig:SchematicFig}
\end{figure}

Compared to the original population dynamics method, the addition of control forces and the use of iteration and feedback increase the complexity of the method presented here.  Here, we summarise the improvements that these changes achieve.  Typically, existing methods either exploit an ensemble (population) of copies of the system~\cite{DMC,Aldous,Grassberger,Lelievre,Garnier,Rolland}, or they use modified (controlled) dynamical rules to drive the system towards rare events of interest~\cite{Nemoto_Sasa_PRL,Berg_Neuhaus1, Wang_Landau1, OrtizLaelbling,DupuisWang,CappeDoucGuillin,ChanLai,hartmann12}, or they use path-sampling methods~\cite{Hedges,speck12}.  All these methods are useful, but the population-based methods can suffer convergence problems, due to the very large populations required in some problems.  On the other hand, the controlled methods require accurate estimation of an optimal control force that is typically a high-dimensional and complex object, which can be difficult to represent computationally (see for example~\cite{JackSollich2}).  Path sampling methods are most effective when the ensemble $P_h$ has time-reversal symmetry, which limits their applicability in non-equilibrium settings. The method proposed here is a mixture of the population-based and control-based methods, as illustrated schematically in Fig~\ref{fig:SchematicFig}.

In terms of the applicability of this new method, we expect the following general behaviour.  For complex high-dimensional problems, accurate representation of the optimal control $V^*$ is likely to be difficult, but we expect even approximate representations of $V^*$ to significantly improve the performance of the population dynamics method.  Thus, the controlled method should reduce the computational cost of problems that are already tractable using population dynamics, allowing access (for example) to larger system sizes and larger values of the bias parameter $h$.  On the other hand, for relatively simple problems such as the particle in a quartic potential of Sec.~\ref{Subsec:NumericalExample}, the original population dynamics fails for small noise (Fig.~\ref{kappa01ComputationalEfficiencyFeedback}) but we would expect that a solution by the controlled method of~\cite{Nemoto_Sasa_PRL} might already be possible.  However, for a similar model in three or more dimensions, we expect that the method of~\cite{Nemoto_Sasa_PRL} would already be challenging, due to the difficulty of representing exactly the effective potential.  Here, we combine that control strategy with population dynamics: we arrive at a flexible method that exploits the strengths of both approaches, and which we anticipate will be effective in a wide variety of problems.

\section{Numerical example}
\label{Sec:NumEx}

To illustrate the control-with-feedback method, we consider the numerical example from Section \ref{Subsec:NumericalExample}, and we take the effective description in (\ref{eq:effective_definition}) to be a quartic polynomial:
$\fn_i (x) \equiv x^{i}$ (that is, ``$x$ raised to the power $i$'') and $\cut = 4$.
For the first iteration of the method we take $(c_i^0)_{i=1}^{\cut}=0$.
Note that this potential-parametrisation of $V$ cannot capture the exact $V^*$, neither for $\epsilon >0$ nor in the limit $\epsilon \rightarrow 0$ (see Appendix~\ref{appendix:InstantonResults}).  This emphasises that the control-with-feedback method does not require a perfect representation of the optimal control in order to improve the convergence of the population dynamics method.

Fig.~\ref{kappa01ComputationalEfficiencyFeedback} shows estimates of the distribution $p_{\rm ave}$ obtained using the original cloning method (Fig.~\ref{kappa01ComputationalEfficiencyFeedback}(a-d)), compared with the results obtained
using control-with-feedback procedure proposed here  (Fig.~\ref{kappa01ComputationalEfficiencyFeedback}(e)).  (Two iterations of the feedback were used, which allow an accurate estimate of the optimal control potential $V^*$.)
The comparison between Fig.~\ref{kappa01ComputationalEfficiencyFeedback}(d) and Fig.~\ref{kappa01ComputationalEfficiencyFeedback}(e) shows that the number of clones required to obtain convergence to the exact result 
is much reduced using the control-with-feedback method.

In the weak-noise limit $\epsilon \rightarrow 0$, one can see this advantage more clearly.
In this limit, a sampling issue arises because of the exponential increase of the required number of copies $N_c$,
as discussed in Section~\ref{subsec:SmallNoiseProblem}.
Fig.~\ref{fig:NtGh}(a) shows numerical results for $\tilde G(\tilde h)$, as $\epsilon$ is reduced. 
 The normal population dynamics  converges very poorly for small noise, $\epsilon<\epsilon^*$. However, the controlled population dynamics does not fail at small $\epsilon$ because it maintains $p^w_{\rm end} \approx p_{\text{ave}}^w$~\cite{footnotecitation_anneaing}.
%Note that the quartic approximation %of $V$ for $w^*$
%considerably reduces the systematic errors, \emph{even though it does not capture the exact $w^*$}~\cite{SM}.

%% Statistical error
We then consider \emph{statistical} errors.  Fig.~\ref{fig:NtGh}(b) shows 
the number of distinct clone positions in the population, $\tilde{N}_c(t)$. 
Again, the control-with-feedback method performs better than the original method, in that it averages over a larger sample of distinct positions, reducing the statistical errors.  
%The method yields the function $G(h)$, which specifies the probabilities of rare events; it also generates sample paths from $P_h(X)$.  Finally, the optimal control provides a simple \emph{physical} strategy for promoting the rare events of interest in this illustrative model system.

\begin{figure}
\centering
%\subfigure[~$p_{\rm end}^w(x) \left (  \left (\frac{p_{\rm ave}^w(x)}{p_{\rm end}^w(x)} \right )^2 - 1  \right ) $]{
\includegraphics[width=0.9\columnwidth]{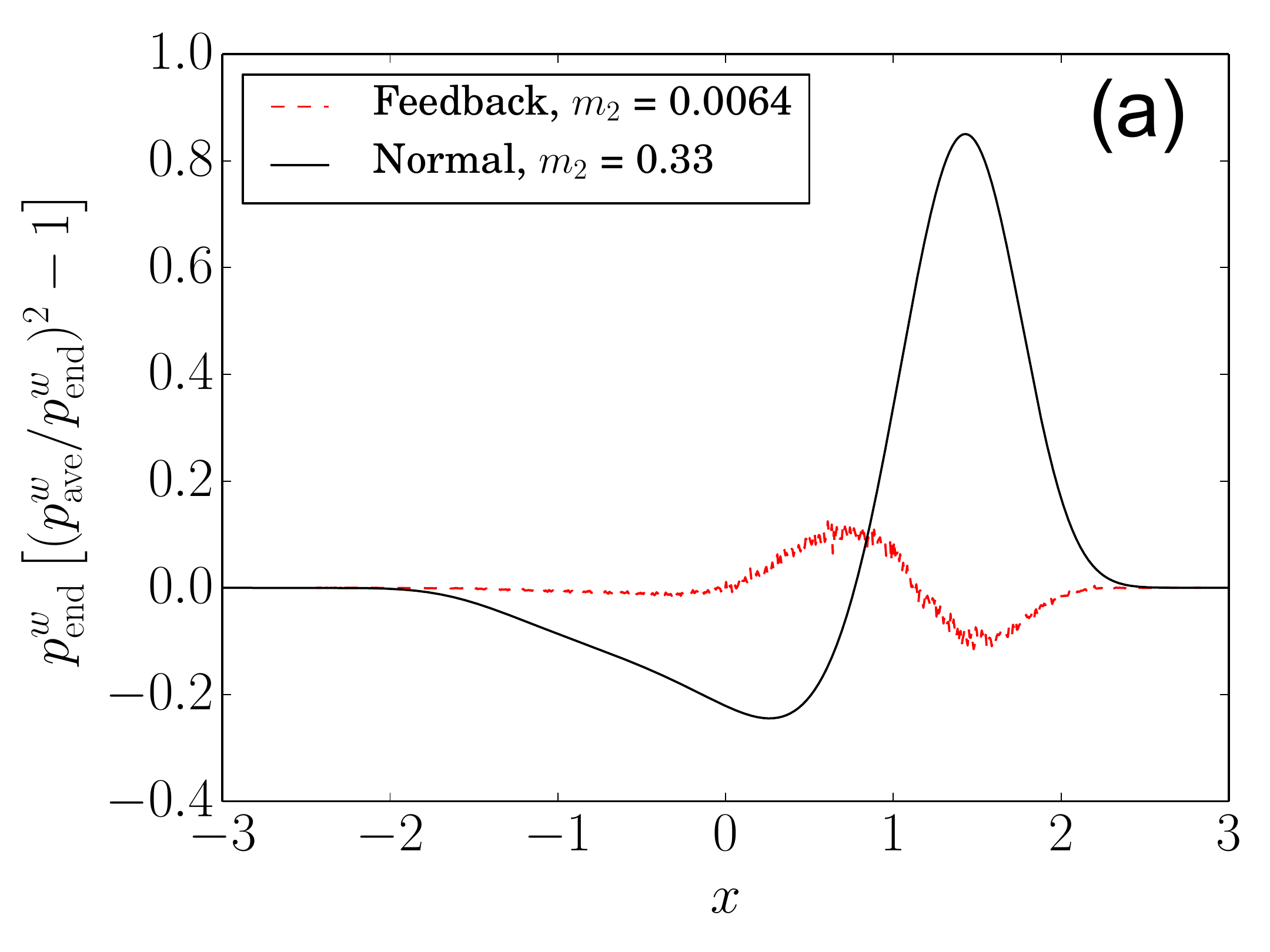} \\%}
%\subfigure[~$p_{\rm ave}^w(x) \log \left (\frac{p_{\rm ave}^w(x)}{p_{\rm end}^w(x)} \right )$]{
\includegraphics[width=0.9\columnwidth]{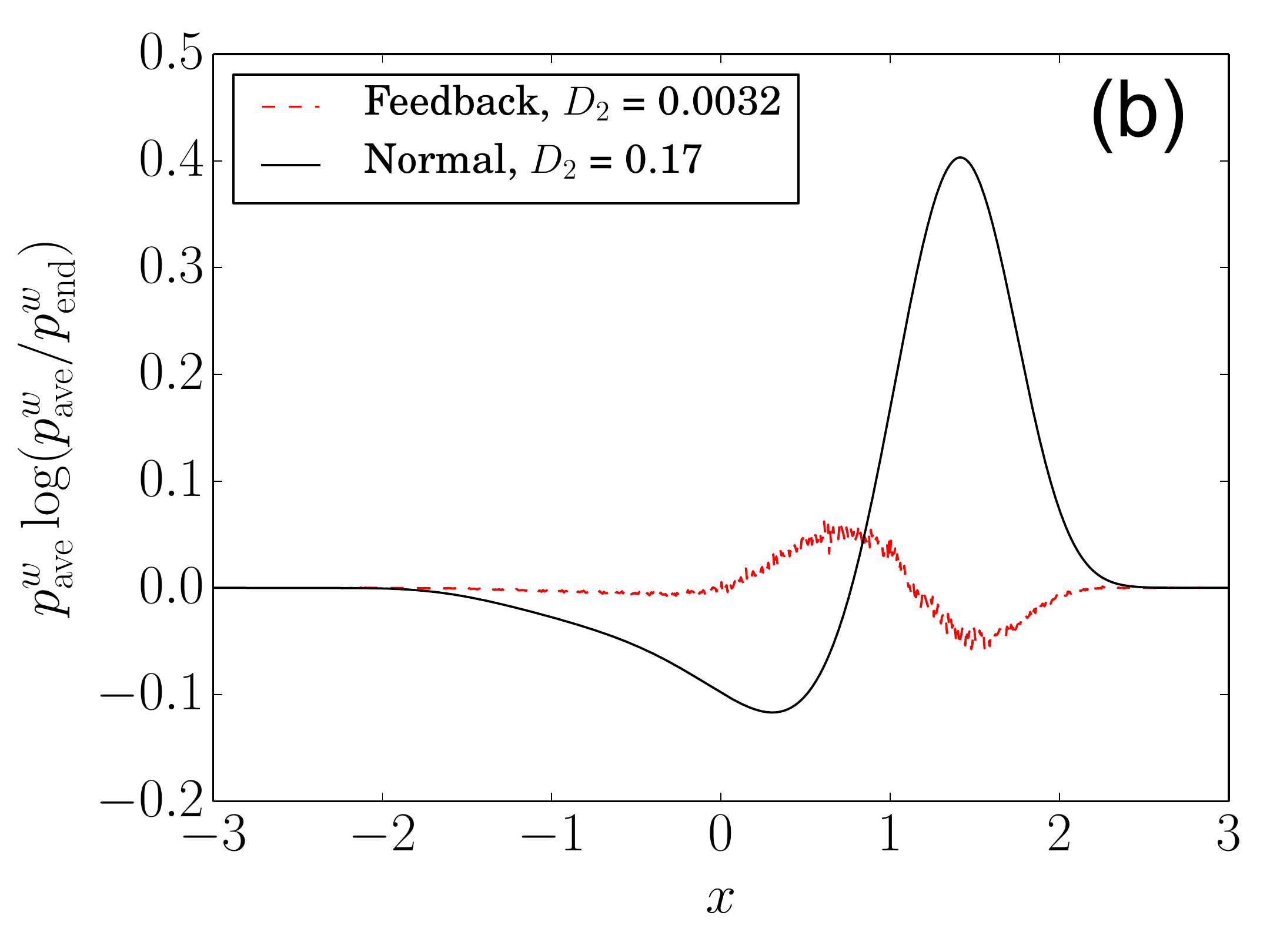}%}
\caption{
(Color online) The integrands of $m_2$ and $D_2$ defined as (a) $p_{\rm end}^w(x) \left [ \left (\frac{p_{\rm ave}^w(x)}{p_{\rm end}^w(x)} \right )^2 -1  \right ]$ and (b) $p_{\rm ave}^w(x) \log \left (\frac{p_{\rm ave}^w(x)}{p_{\rm end}^w(x)} \right )$, (see (\ref{equ:m2def}) and (\ref{equ:d2def})),  for the standard (normal) population dynamics method ($w=0$) and for the control-with-feedback method ($w$: obtained from the control-with-feedback method)). For the control-with-feedback method, we set $N_c=20$. In the legends of the figures, we put the values corresponding to $m_2$ and $D_2$.}
\label{fig:multiplicitiesDiscussions}
\end{figure}

%% Multiplicities discussions
Finally, in order to illustrate how the control-with-feedback method improves the standard population dynamics method, in Fig.~\ref{fig:multiplicitiesDiscussions}, we show the integrands of $m_2$ and $D_2$ defined in (\ref{equ:m2def}) and (\ref{equ:d2def}) \cite{footnote:m2andD2}.
As discussed in Section~\ref{subsec:Multiplicity} and Section~\ref{subsec:SmallNoiseProblem}, the standard population dynamics has sampling issues, which are captured by the deviations of $m_2$ and $D_2$ from 0. In the figure, we can see that the control-with-feedback method greatly reduces the values of $m_2$ and $D_2$ close to 0, ensuring that $p_{\rm end}^w$ and $p_{\rm ave}$ are closer than in the original cloning, thus yielding better performances as seen throughout this section.

\section{Outlook} 
\label{Sec:outlook}
We have shown that the performance of the population dynamics algorithm for sampling large deviations~\cite{Populationdynamics} can be improved by introducing a controlling force $w$.  Given the optimal choice for this force, the rare events of interest in large deviation theory
can be characterised as typical trajectories of the controlled system without any cloning. 
In complex systems with many degrees of freedom it is likely that the optimal $w$ cannot be determined exactly, but even non-optimal controls can still
significantly improve both the statistical and the systematic errors associated with the population dynamics method  
(see Section~\ref{Sec:NumEx}).  
It is straightforward to
improve existing population dynamics codes to include this approach: we expect that it will significantly expand the range of systems for which
numerical calculations can be performed, including open quantum systems~\cite{Garrahanquantum,QuantumGarrahan}, or more complex molecular dynamics models than
those considered so far~\cite{Hedges,speck12}. 
%Furthermore, since this method combines the new idea into population dynamics, it might shed light on well-known challenging problems, such as the calculation of concave large deviation functions \cite{Touchette}.

\begin{acknowledgments}
The authors gratefully acknowledge the  support of Fondation Sciences Math\'ematiques de Paris -- EOTP NEMOT15RPO, PEPS LABS and LAABS Inphyniti CNRS project.
The research leading to these results has received funding from the European Research Council under the European Union's seventh Framework Programme (FP7/2007-2013 Grant Agreement No.~616811) (F.~Bouchet and T.~Nemoto).
V.~Lecomte acknowledges support by the National Science Foundation under Grant No.~NSF PHY11-25915 during a stay at KITP, USCB.

\end{acknowledgments}

\appendix

\section{Population dynamics method}
\label{PDSection}
In this appendix, complementing Section~\ref{Subsec:PopulationDynamics} and Section~\ref{Subsec:pend_pave},
we explain the details of the population dynamics algorithm.

\subsection{Population dynamics algorithm}
\label{PDSection_Algorithm}
The population dynamics is a numerical technique designed to evaluate a large deviation function associated to the cumulant generating function (CGF) of a time-averaged observable~$\Lambda(t)$. 
Each step of the algorithm consists of a first sub-step in which the normal (unbiased) dynamics of the system is simulated for a time $\Delta T$, followed by an elimination-multiplication sub-step.  (The elimination-multiplication sub-step is also called a cloning step, or a mutation-selection step.)  
In detail, the method is:
\begin{enumerate}
\item Generate $N_c$ initial conditions, for example, drawn from the stationary state of the unbiased ($h=0$) dynamics.  
 
 \item Repeat the following procedure $M$ times. (The iteration index is $m=0,1,\dots,M-1$.)

	\begin{enumerate} 
	\item For each copy of the system, perform the normal dynamics from $t = m\, \Delta T$ to $(m+1)\Delta T$. We denote each trajectory by $x^a(t)$. (Throughout this section, $a=1,2,....,N_c$.) During the simulation, for each trajectory, calculate 
	\begin{equation}
	s_a = \exp \Big \{ h \left [ (t+\Delta T)\,\Lambda(t+\Delta T) - t \Lambda(t) \right ]\Big \}.
	\end{equation}

	\item For each trajectory $a$, calculate an integer $n_a$ as
	\begin{equation}
		n_a =\left \lfloor \frac{s_a}{\sum_{b} s_b} N_c + \eta \right  \rfloor,
	\end{equation}
	where $\eta$ is a random number uniformly distributed on [0,1] and $\lfloor\cdot\rfloor$ denotes the lower integer part. 
	Calculate and store the quantity $S_m = \sum_{b} s_b$. 
	
	\item Multiply or eliminate each trajectory $a$ so that it appears $n_a$ times in the new population. (For example, if $n_a=0$ then trajectory $a$ is deleted. If $n_a=5$ then we retain trajectory $a$ and we introduce 4 new copies of that trajectory.) 
	
	\item Eliminate or multiply trajectories within the population, chosen randomly and uniformly, so that the total number of surviving trajectories is $N_c$.
	
	\item Go back to (a), using the current set of configurations $x^a((m+1)\Delta T)$ as initial conditions for the next iteration of the normal dynamics.

	\end{enumerate}

\end{enumerate}

Note that if the population were not kept constant in step 2c above, then the population would expand by a factor of $S_m/N_c$.  It follows that the CGF %\cite{footnote:exponentialgrowth} 
can be estimated as
\begin{equation}
G(h) \simeq \frac{1}{M\,\Delta T}\sum _{m} \log \frac{S_m}{N_c}.
\end{equation}
Also, averages over the population at the final time $\tau$ are estimates of averages with respect to $p_{\rm end}$: 
\begin{equation}
\int f(x) p_{\rm end}(x) \dd x \simeq \frac{1}{N_{\rm c}} \sum_{a=1}^{N_{\rm c}} f(x^a(\tau)),
\label{supp_pend}
\end{equation}
which follows from the definition of $p_{\rm end}$.
%where these formulas are directly obtained from their definitions. 
When estimating $p_{\rm end}$, we can improve the statistics by using the history of $x^a(t)$. That is, assuming an ergodicity property, we can replace $f(x^a(\tau)))$ by its time average, leading to
\begin{equation}
\int f(x) p_{\rm end}(x) \dd x \simeq \frac{1}{\tau N_{\rm c}}  \int_{0}^{\tau} \sum_{a=1}^{N_{\rm c}}  f(x^{a}(t)) \dd t .
\end{equation}
This means that the empirical distribution of $x_a(t)$ is an estimator for $p_{\rm end}$, as announced in (\ref{eq:pendestimation}).

In order to generate the sample paths corresponding to the biased measure $P_h$, we also need to copy the history of trajectory (not just the current configuration of $x$) in the selection-mutation procedure in step 2.(b) of the algorithm. This fact is directly derived from the definition of $P_h$. Thus, the $x^a(t)$ defined above do not correspond to sample paths of $P_h$.  The paths are obtained as $\tilde x^a(t)$, which are defined as those trajectories that survive until the final time $\tau$ (see Fig.~\ref{fig:traj}).  In numerical simulations, there are several ways to generate (or reconstruct) these trajectories, as we now explain. 

\subsection{Generating continuous sample paths $\tilde x^a(t)$ for the biased dynamics}
\label{Auxiliarypopulationdynamics}

A simple way to characterise $\tilde x^a(t)$ is the following: If we do not require full sample paths but only wish to evaluate the biased average of an additive observable $A(\tau)=\int_0^\tau  a(x(t))\,\dd t$, a simple method~\cite{Julien_Vivien}
consists in attaching a value of the observable~$A$ to every trajectory and, at every time step, to update its value and copy/delete it together with the trajectory. Then, an evaluation of the biased average of~$A$ is given by an average of the numerical values of~$A$: this average runs over all trajectories that are present {\it at the final time}.
%The issue with this method is that it depends explicitly on the observable~$A$. 
For example, when we divide the configuration space into small bins and take $a_i(x(t))=1$ if $x(t)$ is in bin $i$, $A_i(\tau)/\tau$ is an estimate of $p_{\rm ave}$, integrated across the $i$th bin. %(See Section~\ref{Sec:NumEx} for example.)

For the small systems where we can store all of the trajectories in the population dynamics, we can 
generate full sample paths corresponding to $\tilde x^a$. 
The procedure is as follows:
we first generate all the trajectories, and then select those that survive until the final time $\tau$.
Considering the $N_c$ copies at final time, indexed by $1\leq a\leq N_c$, one can follow the ancestors of every copy. Upon every coalescence observed backwards in time (corresponding to multiplications of clones in the original forwards simulation), one increments a counter $m_a(t,\tau)$ by the number of trajectories which have coalesced. At the end of the procedure, the counters $(m_a(t,\tau))_{1\leq a\leq N_c}$ represent, at time $t$, the number of descendants of a copy $a$ at final time $\tau$.

\section{Derivation of the ratio of path probability density (\ref{equ:phw})}
\label{Sec:App_Derivation}
In this appendix, complementing Section~\ref{Sec:FB_PD}
we derive the relation between $P_{h}[X]$ and $P_{w}[X]$, eq.(\ref{equ:phw}).
We show the derivation in two ways, one based on path probability densities (stochastic differential equations) and the other on Fokker-Plank equations. 

\subsection{Derivation using path probability density}

We denote a trajectory of the system by $X=(x(t))_{0\leq t\leq\tau}$.  From the definitions of $P_{h}[X]$ and $P_{w}[X]$, we have
%\begin{equation}
\begin{multline}
\frac{P_{w}[X]\ee^{-h\tau \Lambda^w(\tau)}}{P_{h}[X]\ee^{-h\tau \Lambda(\tau)}} 
 =  \exp \Bigg [  \int_{0}^{\tau}  \left ( \dot x - F \right ) \cdot \kappa^{-1}    w \dd t  \\ - \frac{1}{2}\int_{0}^{\tau}   w \cdot \kappa^{-1}     w \dd t \Bigg ]
\label{supp:eq:9}
\end{multline}
%\end{equation}
The integrand on the right-hand side is written as
%\begin{equation}
\begin{multline}
 \left ( \dot x - F \right ) \cdot  \kappa^{-1}  w
- \frac{1}{2} w\cdot \kappa^{-1} w \\
 =   \dot x\cdot  \left ( - \nabla V + h \lambda_{\rm c} \right )
- \left ( F +  \frac{1}{2} w \right )\cdot \kappa^{-1}  w
\label{supp:eq:10},
\end{multline}
%\end{equation}
where we have used the expression of $w(x)$ as given in the main text ($w(x) = \kappa \left [ - \nabla V(x) + h \lambda_{\rm c}(x) \right ]$). We then consider the integral of the first term on the right hand side:
\begin{equation}
\int _{0}^{\tau}  \dot x \cdot  \left ( - \nabla V \right ) \dd t.
\label{supp:eq:11}
\end{equation}
Since the trajectory $X$ is generated from the stochastic differential equation (\ref{modifiedsystemdef}) and we use the It\=o convention, the time-derivative of $V(x(t))$ is given by It\=o's formula
\begin{equation}
\frac{\dd }{\dd t}V = \dot x \cdot  \nabla V + \frac{1}{2}{\rm Tr}\left [ B^T  H_V B \right ].
\label{supp:eq:12}
\end{equation}
Here $H_{V}$ is a Hessian matrix defined as $(H_{V})_{i,j}=\frac{\partial V}{\partial x_i \partial x_j}$. 
%In this formula, we assume that the diffusion term in $\dot x$ is described by 
%$B(x(t))\xi(t)$ as the original Langevin equation [eq.(1) in the main text]. 
Combining (\ref{supp:eq:12},\ref{supp:eq:11}) we have
%\begin{equation}
\begin{multline}
\int _{0}^{\tau}  \dot x \cdot \left ( - \nabla V \right ) \dd t 
=  - V(x(\tau))
+ V(x(0)) \\ +  \int _{0}^{\tau}  \frac{1}{2}{\rm Tr}\left [ B^T  H_V B \right ] \dd t.
\label{supp:eq:13}
\end{multline}
%\end{equation}
Thus, from (\ref{supp:eq:9}), (\ref{supp:eq:10}) and (\ref{supp:eq:13}), 
we get
\begin{multline}
%\begin{split}
\frac{P_{w}[X]\ee^{-h\tau \Lambda^w(\tau)}}{P_{h}[X]\ee^{-h\tau \Lambda(\tau)}} 
%\\
=  \ee^{-V(x(\tau))+V(x(0))}  
\\ \times \exp \Bigg \{  \int_{0}^{\tau}  \Big [ \frac{1}{2}{\rm Tr}\left [ B^T  H_V B \right ] 
%\\
 %\qquad \qquad \qquad  \qquad 
 +  h \dot x \cdot    \lambda_{\rm c} \\
- \left ( F +  \frac{w}{2}  \right ) \cdot \kappa^{-1}   w \Big ] \dd t \Bigg \}
\label{supp:eq:14}
%\end{split}
\end{multline}
Finally, by noticing ${\rm Tr}\left [ B^T  H_V B \right ] = {\rm Tr}\left [ H_V \kappa \right ] $ and using the definitions of $\Lambda^ w$ and $\Lambda$, the right hand side is
$\ee^{-V(x(\tau))+V(x(0))} \ee^{h\tau \Lambda(\tau) - h \tau \Lambda^w(\tau)}.$
Hence one arrives at Eq.~(\ref{equ:phw}).

\subsection{Derivation using time-evolution operator}
\label{sec:fp}

An alternative derivation of (\ref{equ:phw}) is obtained by using a `tilted' generator (or master operator) for the biased ensemble of trajectories.  
Let $u^h(x,\tau)$ be the (unnormalised) probability density at time $\tau$, obtained as a marginal of the path distribution $P_h$.  As discussed, for example, in Appendix A.2 of \cite{Chetrite_Touchette2}, this distribution evolves in time  according to a generalised Feynman-Kac formula as 
\begin{equation}
\frac{\partial}{\partial \tau} u^h = L^h[u^h],
\label{supp:eq:18}
\end{equation}
with
\begin{equation}
\begin{split}
L^h[f]  \equiv &  L_{\rm FP}^{F}[f] +  h \left ( \lambda_{\rm d} + \lambda_{\rm c}\cdot F \right ) f \\
&+ \frac{h^2}{2} \left ( \lambda_{\rm c}\cdot \kappa \lambda_{\rm c} \right )  f- h \nabla \cdot \left ( \kappa  \lambda_{\rm c} f \right ) .
\label{equ:lh}
\end{split}
\end{equation}
Here, the Fokker-Planck operator $L_{\rm FP}^{F}$ is
\begin{equation}
L_{\rm FP}^{F}[f] = - \nabla \cdot \left [   F f \right ] + \frac{1}{2} \sum_{i,j} \frac{\partial ^2}{\partial x_i\partial x_j} \kappa_{ij} f,
\label{supp:eq:21}
\end{equation}
where the superscript $F$ on $L_{\rm FP}^{F}$ indicates that the particle feels the physical force $F$ introduced in (\ref{equ:dx}).
 
For the controlled population dynamics, the analogue of $u^h$ is
$u^{w}(x,\tau)$, which evolves as $\frac{\partial}{\partial \tau} u^{w} = L^{w}[u^w]$, with 
\begin{equation}
L^w[f]  \equiv  L_{\rm FP}^{F+w}[f] +  h \lambda^w f
\label{equ:lw}.
\end{equation}

The relation (\ref{equ:phw}) follows from a duality relation between $L^h$ and $L^w$:
\begin{equation}
L^h[f]= \ee^{V} L^w[f \ee^{-V}].
\label{supp:eq:21_1}
\end{equation}
This relation may be verified directly from (\ref{equ:lh},\ref{equ:lw}), noting that the potential $V$ is related to the control $w$ via the definition $w=h\kappa\lambda_{\rm c} - \kappa\nabla V$.  

From (\ref{supp:eq:18}), we note that the operator $U^h_\tau=\ee^{\tau L^h}$ corresponds to integration forward in time over a duration $\tau$.  Similarly $U^w_\tau={\rm e}^{\tau L^w}$, and from (\ref{supp:eq:21_1}) we have
$U^h_\tau[f] = \ee^{V} U^w_\tau[f \ee^{-V}]$.  Setting $f(x)=\delta(x-x_0)$, then $u^h(x,\tau|x_0,0)=U^h_\tau[f]$ is the (unnormalised) probability density at $x$, for a particle that was at $x_0$ a time $\tau$ earlier.  Defining similarly $u^w(x,\tau|x_0,0)$, (\ref{supp:eq:21_1}) implies 
\begin{equation}
u^h(x,\tau|x_0,0) = \ee^{V(x)} u^w(x,\tau|x_0,0) \ee^{- V(x_0)} .
\end{equation}
Hence one arrives at (\ref{equ:phw}) of the main text.

This approach also provides insight into the distributions $p_{\rm ave}$ and $p_{\rm end}$, as discussed in~\cite{Populationdynamics,JackSollich}.  One easily sees that 
\begin{equation}
p_{\rm end}(x) = \lim_{\tau\to\infty} \frac{u^h(x,\tau|x_0,0)}{\int_{x'} u^h(x',\tau|x_0,0)},
\label{equ:pend-ev}
\end{equation} 
which is independent of $x_0$.  
Similarly, 
\begin{equation}
p_{\rm ave}(x) = \lim_{\tau\to\infty} \frac{ \int_{x_1} u^h(x_1,\tau/2|x) u^h(x,0|x_0,-\tau/2)}
     {\int_{x',x_1} u^h(x_1,\tau/2|x') u^h(x',0|x_0,-\tau/2)}.
\end{equation} 
For large $\tau$, the propagator $u^h$ is dominated by the largest eigenvalue of $L^h$, as
\begin{equation}
u^h(x,\tau|x_0,0) \simeq p_{\rm end}(x) {\rm e}^{G(h)\tau} q(x_0),
\label{equ:eh-ev}
\end{equation}
 where $p_{\rm end}(x)$ is the dominant right eigenvector of $L^h$ (required for consistency with (\ref{equ:pend-ev})), the associated eigenvalue is $G(h)$, and $q(x)$ is the dominant left eigenvector.  The approximate equality in (\ref{equ:eh-ev}) is valid for large times, up to corrections of order ${\rm e}^{-\lambda \tau}$, where $\lambda$ is the spectral gap of $L^h$. Combining (\ref{equ:pend-ev}-\ref{equ:eh-ev}) we have $p_{\rm ave}(x) \propto p_{\rm end}(x) q(x)$.

This approach also shows why $p_{\rm ave}$ is not affected by the control force $w$: the dominant left and right eigenvectors of $L^h$ are $q$ and $p_{\rm end}$ so (\ref{supp:eq:21_1}) means that the dominant eigenvectors of $L^w$ are $q^w=q{\rm e}^V$ and $p_{\rm end}^w={\rm e}^{-V}p_{\rm end}$.  Hence it is clear that $p_{\rm ave}^w = q^w p_{\rm end}^w  = q p_{\rm end} = p_{\rm ave}$.

%This operator approach also allows to see why the control force~$w$ does not affect the intermediate-time distribution. One notes that 
%$L^w[p^w_\text{end}] = G(h)\, p^w_\text{end}$ (the eigenvalue being $G(h)$ as seen from~\eqref{supp:eq:21_1})
%and that $p^w_\text{ave}=q^w\,p^w_\text{end}$
%where $(L^w)^\dagger[q^w]=G(h)\,q^w$.
%%
%The relation~\eqref{supp:eq:21_1} also provides the connection between the controlled eigenvectors and the original ones at $w=0$: one has $p^w_\text{end}=\ee^{-V}p_\text{end}$ and $q^w=\ee^{V}q$. This implies as announced in the main text that $p^w_\text{ave}=q^w\,p^w_\text{end}=p_\text{ave}$. The end-time distribution $p^w_\text{end}$ is on the contrary affected by the control force~$w$.

In the special case where $w$ is given by the optimal control $w^*$ (that is defined as the control $w$ satisfying the condition $p_{\rm ave}^{w}=p_{\rm end}^{w}$ in the main text), one can show that the controlled system is described by the auxiliary process~\cite{JackSollich} (or the ``driven process''~\cite{Chetrite_Touchette2}), which is a Markov process whose path probability density is equivalent to $P_h$ in its stationary regime. (Indeed, $p_{\rm ave}^{w^*}=p_{\rm end}^{w^*}$ implies $q^{w^*}=1$, which expresses that $L^{w^*}$ conserves probability.)  In this case, one has~\cite{Chetrite_Touchette2}
\begin{equation}
\ee^{-V} L^h[f \ee^V]=  L_{\rm FP}^{F+w^*}[f] + G(h) f,
\end{equation}
where $G(h)$ is a constant (independent of $x$): this is the cumulant generating function. 
Comparing with (\ref{supp:eq:21_1}) one sees that $\lambda^{w^*}(x)$ is independent of $x$, from which it follows that the population dynamics in this case has no cloning or deletion of clones (this property is true for all finite $N_c$: all clones have equal weights at all times).

\section{An example of the feedback-algorithm}
\label{FeedbackPopulationDynamics}

Here, in order to complement Section~\ref{Subsec:IteAFeed}, we explain the algorithm used within the feedback population dynamics.
The procedure is a combination of the population dynamics and an iterative construction of a control potential $V(x)$ that is close to the optimal control $V^*$.  There is considerable flexibility in the precise definitions of the estimators used in this algorithm, but these choices have proven effective in the simple model problem considered here.

\begin{enumerate}
\item Generate $N_c$ initial conditions, for example, drawn from the stationary distribution of the original (unbiased) system. 
%$V(x)$ [or the corresponding parameters $c_i^{w}$ in an effective description (\ref{eq:effective_definition})] is set to be 0 at the beginning.

\item  Repeat the following feedback procedure $R$ times (the iteration index is $r=0,1,\dots,R-1$). We denote by $V^{r}(x)$ the control potential $V(x)$ for iteration $r$ and we take $V^{0}(x) = 0$.

\begin{enumerate}
  \item Perform the population dynamics for the system as explained in Appendix~\ref{PDSection}, using a time interval $M\tau_0$.  The unbiased evolution within the method includes the control force $w^r$ that is obtained from the control potential $V^r$, and the elimination-multiplication step uses the corresponding biasing factor $\Lambda^{w^r}$.  The time $\tau_0$ between elimination-multiplication steps should be larger than the correlation time of the system.  From each time segment (indexed by $m$), estimate the distributions
  \begin{equation}
  p_1^{m,r}(x) = \frac{1}{N_c\tau_0} \sum_{a=1}^{N_c} \int_{m\tau_0}^{(m+1)\tau_0}  \delta[x-x^a(t)] \mathrm{d}t
  \end{equation}
   and 
   \begin{multline}
   p_0^{m,r}(x) = \frac{1}{N_c(\tau_0-t_{\rm end})} \\ \times \sum_{a=1}^{N_c}  \int_{m\tau_0}^{(m+1)\tau_0 - t_{\rm end}}\!\!  \delta[x-{\tilde x}^a(t)] \mathrm{d}t,
   \end{multline}
    where the trajectories $\xt$ are defined on the time interval $[m\tau_0,(m+1)\tau_0]$, as specified in Appendix~\ref{Auxiliarypopulationdynamics}.
  The shift parameter $t_{\rm end}$ is chosen so that $p_0$ is an accurate estimator for $p_{\rm ave}$, by excluding times $t$ that are too close to the final time $(m+1)\tau_0$.  If $\tau_0$ is large enough, all results should depend weakly on $t_{\rm end}$.
  
 %This yields trajectories $(x_{a}^{V}(t))_{a=1}^{N_c}$ and $(\tilde x^{V}_{a}(t))_{a=1}^{N_c}$, as described in Appendix\ref{Auxiliarypopulationdynamics}. %  (with $t\in [0, M \tau_0$]).
%
%
%	\item For a fixed time interval $\tau_0$ (which is sufficiently long compared with the correlation time of the observable $\lambda_{\rm d,c}$), the population dynamics is performed, as explained in Appendix~\ref{PDSection}. From the result, we obtain the trajectories $(x_{a}^{V}(t))_{a=1}^{N_c}$ ($t\in [0, \tau_0]$) and also the ones that have survived on $[0,\tau_0]$
%$(\tilde x^{V}_{a}(t))_{a=1}^{N_c}$ ($t\in [0,\tau_0]$). For the application to large systems, we define a small ({\it local}) phase space, and measure the empirical distribution within this phase space. See also Appendix~\ref{Auxiliarypopulationdynamics}. (Below we do not mention about this local phase space.)
%	
%	
%	\item Then, we start from the final condition of $x_{a}^{V}(t)$ and perform another population dynamics. This allows us to have again the trajectories $(x_{a}^{V}(t))_{a=1}^{N_c}$ ($t\in [\tau_0, 2 \tau_0]$) and also the ones that have survived at time $2\tau_0$ $(\tilde x^{V}_{a}(t))_{a=1}^{N_c}$ ($t\in [\tau_0, 2 \tau_0]$).
%	
%	
%	\item Repeat this procedure $M$ times. We finally have trajectories $(x_{a}^{V}(t))_{a=1}^{N_c}$ ($t \in [0, M \tau_0]$) and also the trajectories $(\tilde x^{V}_{a}(t))_{a=1}^{N_c}$ ($t\in [0, M \tau_0$]).
	
	\item Having completed $M$ time segments within the population dynamics, evaluate $p_{\rm end}^{w,r}(x)$ and $p^{w,r}_{\rm ave}(x)$
	as	
		\begin{align}
			p^{w,r}_{\rm end}(x) &= \frac{1}{M} \sum_m p_1^{m,r}(x),
			%\simeq \frac{1}{M \tau_0 N_c} \sum_{a=1}^{N_c} \int_{0}^{M \tau_0}   \delta (x^r_a(t) - x) \dd t.
			\label{feedbackPDend}
			\\
%		\end{equation}
%		\begin{equation}
			p^{w,r}_{\rm ave}(x) &= \frac{1}{M} \sum_m p_0^{m,r}(x).
%			\begin{split}
%			p^{w,r}_{\rm ave}(x) \simeq & \frac{1}{M (\tau_0 - t_{\rm end} ) N_c} \\
%			&\times \sum_{m=0}^{M-1} \int_{m \tau_0}^{(m+1) \tau_0 - t_{\rm end}}  \sum_{a=1}^{N_c}   \delta (\tilde x^{r}_a(t) - x) \dd t.
			\label{feedbackPDave}
%			\end{split}
		\end{align}
		
% The reason why, in $p_{\rm ave}^{w}(x)$, we set the upper bound of integration to $(r+1)\tau_0 - t_{\rm end}$ is to avoid a transient regime: By setting
% $t_{\rm end}$ to be much larger than a correlation time, one can make the statistics of the result to be better. (We note that if $\tau_0$ is sufficiently large, this setting is not necessary.)

		\item Finally, from these distribution functions, calculate $V^{r+1}(x)$ in terms of a sum of basis functions, according to Eq.~(\ref{eq:sim_definition}).  In practice, note that it is not necessary to keep track of the full distributions $p_0$ and $p_1$, but only those statistics that are required to solve the minimisation in (\ref{eq:sim_definition}).  Also, it is sometimes convenient to take $V^{r+1}(x) =  V^r(x) ( 1 - \alpha ) + V_{\rm new}(x) \alpha$, where $V_{\rm new}(x)$ is the control potential specified by the right hand side of (\ref{eq:sim_definition}), and $\alpha$ is a parameter (with $0 < \alpha \leq 1$) that acts to suppress large fluctuations in $V$.
		% \cite{footnoteiteration}
%		\begin{equation}
%			V^{r+1}(x) =  V^r(x) +   \log \frac{p^{w}_{\rm ave}(x)}{p^{w}_{\rm end}(x) }.
%			\label{W_k+1_and_k}
%		\end{equation}
		%For the effective descriptions, see (\ref{eq:effectiveiteration}) and Section~\ref{Subsec:IteAFeed}. 
	\end{enumerate}
	\item Go back to step $2$ and perform the next iteration $(r+1)$, with the control potential $V^{r+1}$, and initial conditions for the clones given by their current states $x^a(M\tau_0)$.
\end{enumerate}

\section{Langevin system with quartic potential}
\label{appendix:InstantonResults}
In this final appendix, in order to complement Section~\ref{Sec:NumEx}, we explain the property of the system we considered there: the parameters are given by $d=1$, $F(x)=-x^3$, $B(x) = \sqrt{2\epsilon}$, $\lambda_{\rm d}(x) = \lambda(x) \equiv x+x^2$ and $\lambda_{\rm c}(x) = 0$.
We focus on the small-noise limit $\epsilon\to0$.  
Throughout this section, $h$ corresponds to $\tilde h$ in the main text (see below). 

\begin{figure}
\centering
\includegraphics[width=8cm]{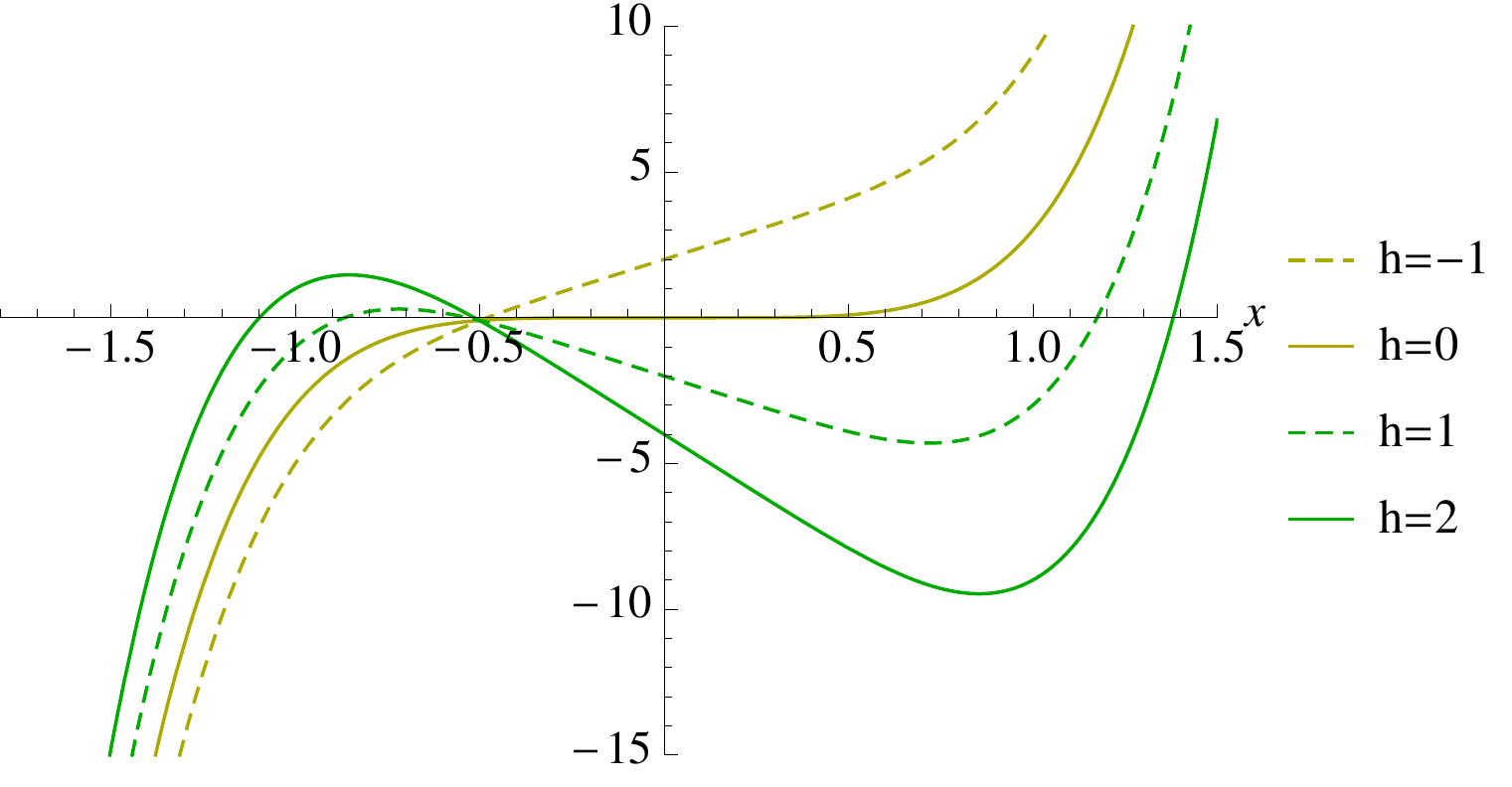}
\caption{\label{StationaryInstanton_quariticpotential} (Color online) Plots of the polynomial $3 x^5 - 4 h x - 2 h$ for several $h$.  The roots
of this polynomial determine the concentration points of $p_{\rm ave}(x)$ for $\epsilon\to0$ in the model system considered below.}
\end{figure}

The main features of the limit $\epsilon\to0$ are
\begin{itemize}
\item The distribution $p_{\rm ave}(x)$ concentrates on a point $x_{\rm ave}$ that is a root of the polynomial
$$
3 x^5 - 4  h x - 2 h =0.
$$
This function is sketched in Fig.~\ref{StationaryInstanton_quariticpotential}.  For $h>0$, the concentration is at the positive root ($x_{\rm ave}>0$);
for $h=0$ one has $x_{\rm ave}=0$.  For negative $h$, the point $x_{\rm ave}$ decreases quickly from zero and localises at $x_{\rm ave}\approx \frac12$. 
\item There is a second-order dynamical phase transition at $h=0$, which appears as divergence of the second derivative of the dynamical free energy, $G''(h)$ (see Fig.~\ref{k0m1a1b1Cumulant}, below).
\item The distribution $p_{\rm end}(x)$ concentrates on a point $x_{\rm end}$, with $x_{\rm end}\neq x_{\rm ave}$ in general.  This leads to poor convergence of the population dynamics method for small $\epsilon$, as discussed in the main text.
\item Even though the system is simple, the analytical expressions of $p_{\rm ave}$ and $p_{\rm end}$ are not straightforward. In particular, the perfect potential $V^*(x)$ corresponding  to $w^*(x)$ is not expressed exactly as the quartic polynomial expansion used to perform a numerical evaluation of $w(x)$ -- however, as described in the main text, this does not affect the effectiveness of the numerical procedure.
\end{itemize}
Below, relying on the Euler-Lagrange equation, we derive the analytical results of $G(h)$, $p_{\rm ave}$ and $p_{\rm end}$ in $\epsilon \rightarrow 0$, from which these features are obtained.

%We show that the system displays a dynamical phase transition. We also derive the analytical expressions of $p_{\rm end}$ and $p_{\rm ave}$ in $\epsilon \rightarrow 0$ limit.

\subsection{Euler-Lagrange equation (Instanton equation)}

We consider the following finite time cumulant generating function:
\begin{equation}
 G_{\tau, \epsilon}(h) = \frac{\epsilon}{\tau} \log \left \langle  \ee^{(\tau/\epsilon) h \Lambda(\tau)}  \right \rangle_{\rm st},
\end{equation}
where $\left \langle \ \right \rangle_{\rm st}$ means the average with respect to the path with a stationary initial condition. (Hereafter, we denote this initial distribution function by $P_{\rm st} (x)$.) The function $G_\epsilon(h)\equiv\lim _{\tau \rightarrow \infty}G_{\tau, \epsilon}(h)$ corresponds to $\tilde G(\tilde h)$ in the main text.
By taking $\epsilon \rightarrow 0$, we obtain the following variational principle:
\begin{equation}
\begin{split}
&\lim_{\epsilon \rightarrow 0}
G_{\tau, \epsilon}(h) \\
& = -\frac{1}{\tau} \min _{x_0,x_\tau}\left [ \min_{\underset{ x(0)=x_0,x(\tau)=x_\tau }{(x(t))_{t=0}^{\tau} } } 
\int_{0}^{\tau}  L(\dot x(t), x(t)) \dd t + F_{\rm free}(x_0) \right ]
\label{variationalprinciple_epsilon0}
\end{split}
\end{equation}
with the Lagrangian $L(\dot x, x)$ defined as 
\begin{equation}
L(\dot x, x) \equiv 
\frac{1}{4}\left (\dot x - F(x) \right )^2  - h \lambda(x),
\end{equation}
and also the free energy function  $F_{\rm free}(x_0)$ defined as
\begin{equation}
F_{\rm free}(x_0) \equiv -  \lim_{\epsilon \rightarrow 0} \epsilon \log P_{\rm st} (x_0) = \frac{1}{4}x_0^4  + \rm const.
\end{equation}
Then, the corresponding Euler-Lagrange equation (Hamilton equation), 
which is obtained from minimising this action, is
\begin{equation}
\dot x  = -x^3 +2p
\label{Lagrange_x_2}
\end{equation}
\begin{equation}
\dot p   = 3  p x^2  - h (2x+1),
\label{Lagrange_P}
\end{equation}
with the required initial and the final conditions as
\begin{equation}
p(0) = \frac{\partial F_{\rm free}(x)}{\partial x} \bigg | _{t=0} = x(0)^3 
\label{generalpinitial}
\end{equation}
\begin{equation}
p(\tau)= 0.
\label{generalpfinal}
\end{equation}
We analyse these equations numerically and analytically in \cite{InPreparation}. The following results
are based on that study.

\subsection{Steady solutions}
Here, we consider the steady solutions of these instantons, which is defined as the solution obtained from $\dot x_{\rm st}=\dot p_{\rm st}=0$ in (\ref{Lagrange_x_2}) and (\ref{Lagrange_P}).
These conditions lead to
\begin{equation}
p_{\rm st}=\frac{1}{2} x_{\rm st}^3 
\end{equation}
and
\begin{equation}
3 x_{\rm st}^5 - 4  h x_{\rm st} - 2 h =0.
\label{xst_determine_quartic2}
\end{equation}
We plot the left-hand side of (\ref{xst_determine_quartic2}) as a function of $x$ in Fig.~\ref{StationaryInstanton_quariticpotential} for several fixed $h$.
The figure shows that this equation has three solutions, when $h$ is larger than a certain value (larger than 0).

 \begin{figure*}%[h]
\centering
%\subfigure[$\lim_{\epsilon \rightarrow 0}G_{\epsilon}(h)$]{
\includegraphics[width=.85\columnwidth]{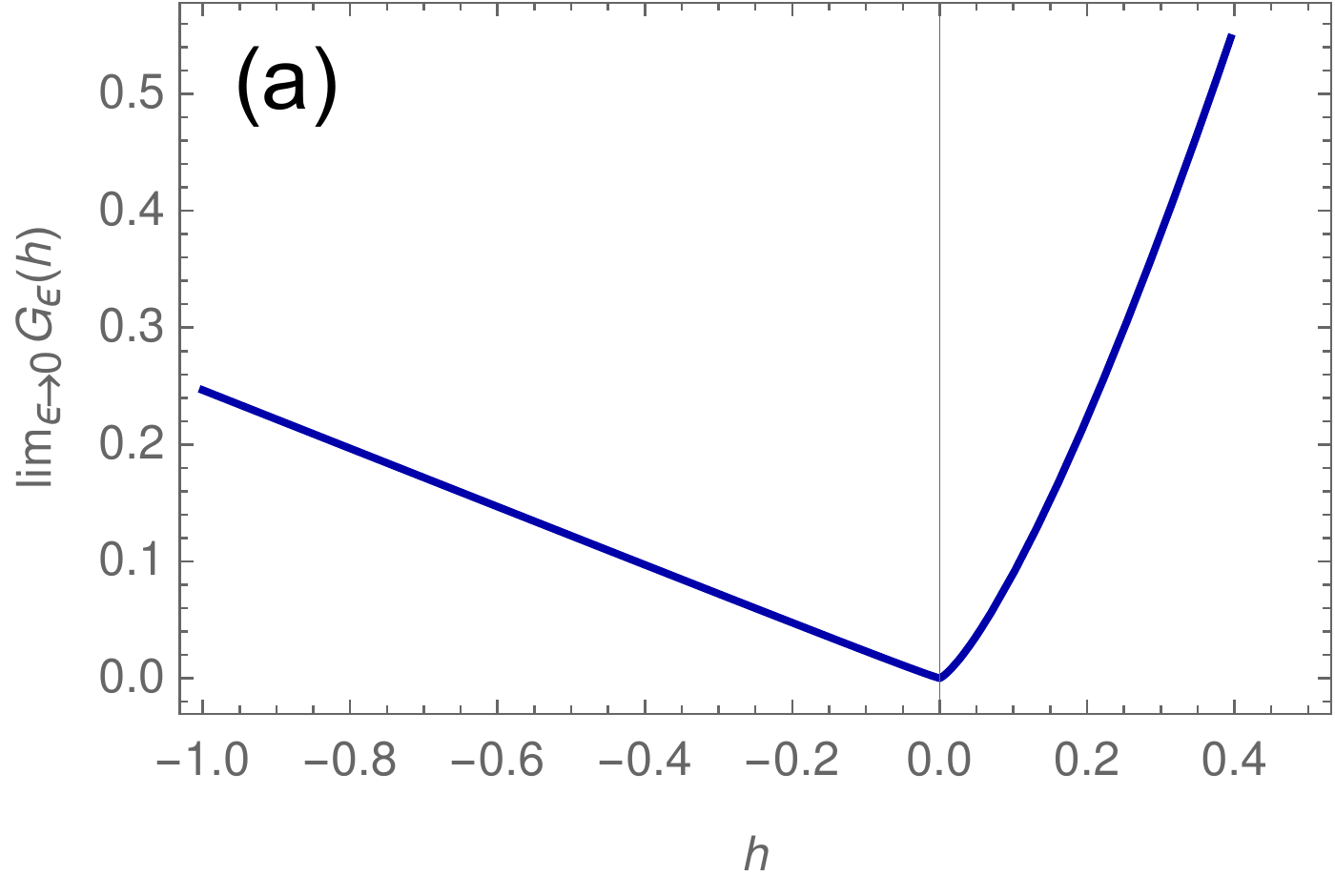}%}
%\subfigure[Green solid line: $(d/dh)\lim_{\epsilon \rightarrow 0}G_{\epsilon}(h)$, Red dotted line: $-(d/dh)\lim_{\epsilon \rightarrow 0}G_{\epsilon}(-h)$, Black solid line: $h^{1/5}$]{
\includegraphics[width=.9\columnwidth]{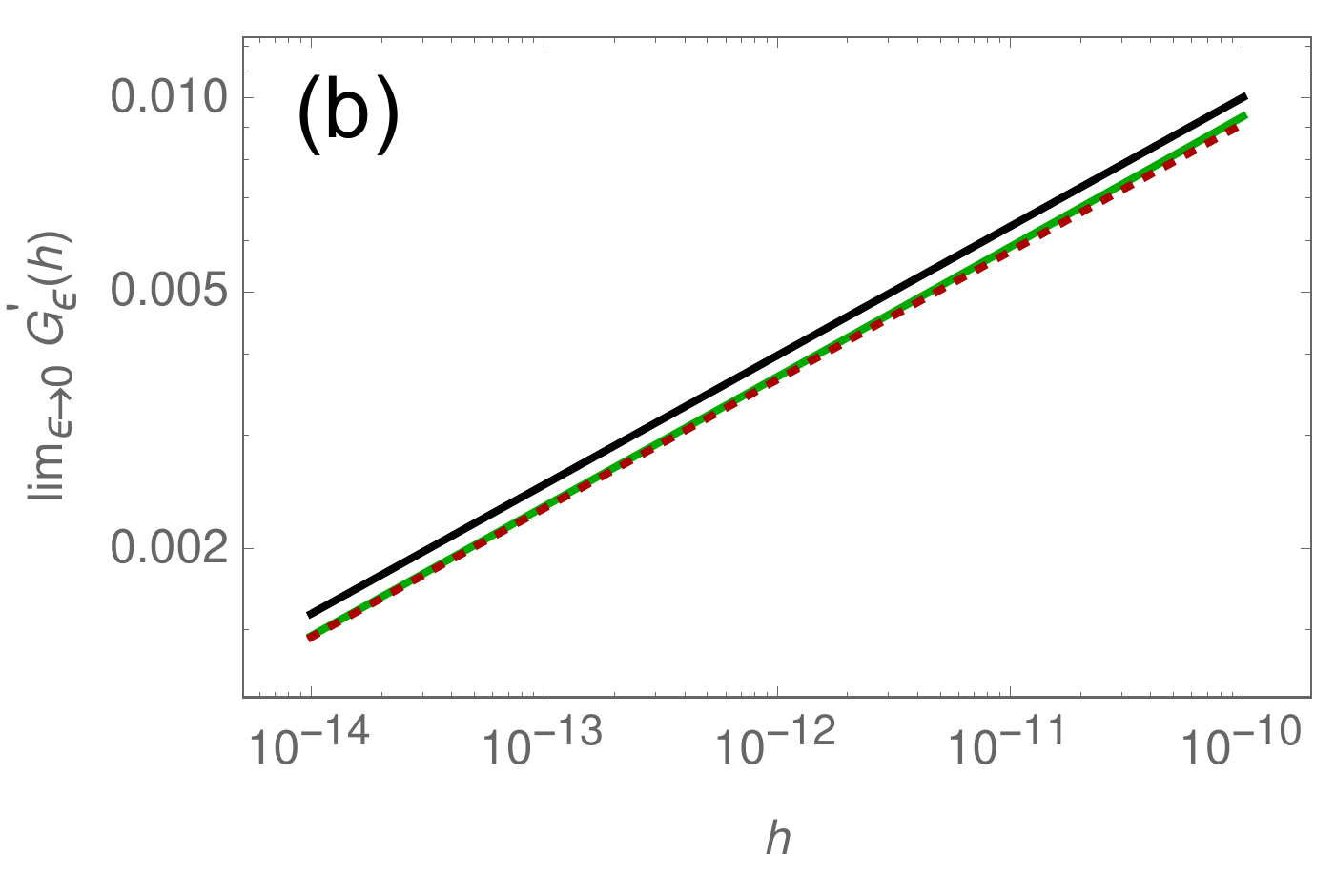} \\%}
%\subfigure[$(d^2/dh^2)\lim_{\epsilon \rightarrow 0}G_{\epsilon}(h)$]{
\includegraphics[width=.9\columnwidth]{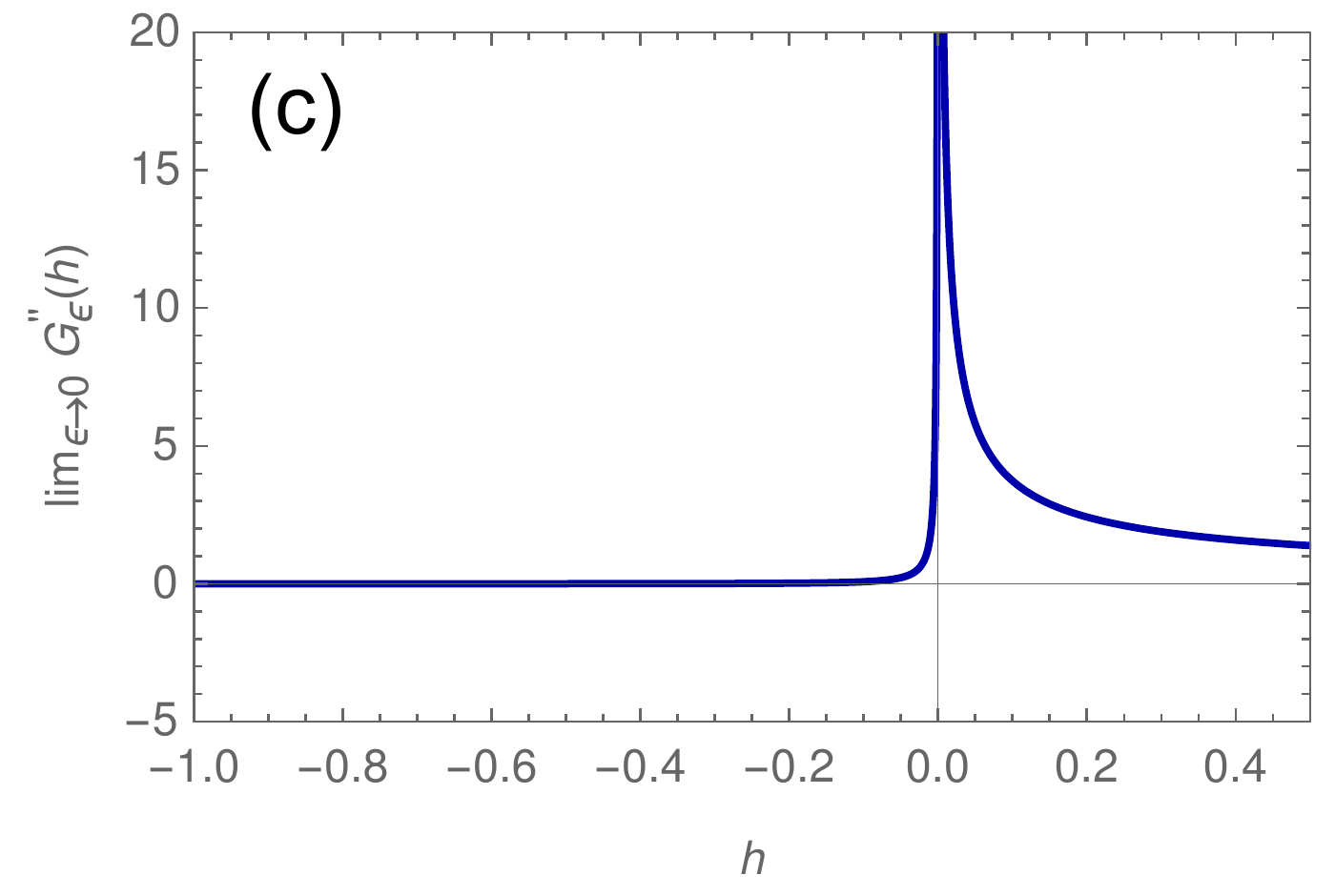}%}
\caption{\label{k0m1a1b1Cumulant} (Color online) 
\textbf{(a)} Generating function $\lim_{\epsilon \rightarrow 0}G_{\epsilon}(h)$. \textbf{(b)} The first derivatives, where the green (dark grey in the printed version) solid line represents $(d/dh)\lim_{\epsilon \rightarrow 0}G_{\epsilon}(h)$, the red dotted line represents $-(d/dh)\lim_{\epsilon \rightarrow 0}G_{\epsilon}(-h)$ and the black solid line represents a straight line $h^{1/5}$. From this panel, we find that the first derivative converges to 0 as a power law $\sim h^{1/5}$, (as can also be checked analytically). \textbf{(c)} The second derivative of $\lim_{\epsilon \rightarrow 0}G_{\epsilon}(h)$ with respect to $h$. These are calculated from~(\ref{steadyassumption}-\ref{Gst}).
We can see that the second derivative shows a singularity at $h=0$, although the first derivative converges to 0. This represents a second-order dynamical phase transition.
}
%%%
%%% from order-dyn-transition_quartic.nb
%%%
\end{figure*}

\subsection{Cumulant generating function}
\label{appendix:subsectionCumulant}
From the variational principle (\ref{variationalprinciple_epsilon0}), even in the case where there are multiple instanton solutions, the cumulant generating function can be calculated. This is based on the observation that 
the instanton solution corresponding to the minimum is time-independent \cite{footnote:linearstabilityanalysis}.
More precisely, by combining this observation with the variational principle (\ref{variationalprinciple_epsilon0}), we get
\begin{equation}
\lim_{\epsilon \rightarrow 0}G_{\epsilon}(h)=\max_{x_{\rm st}}G_{\rm st}(x_{\rm st}),
\label{steadyassumption}
\end{equation}
with
\begin{equation}
G_{\rm st}(x_{\rm st})  \equiv - \frac{1}{4}x_{\rm st}^6 + h \left (  x_{\rm st}^2 +  x_{\rm st} \right ).
\label{Gst}
\end{equation}

We plot the $\epsilon \rightarrow 0$ result, $\lim_{\epsilon \rightarrow 0}G_{\epsilon}(h)$, in Fig.~\ref{k0m1a1b1Cumulant}, from which we can see that the generating function has a kink at the origin, which is the sign of the dynamical phase transition in this system, appearing in the zero-temperature limit.% \cite{footnote:secondorder}.
Asymptotic analysis allows to find $G(h)\sim A_{\pm}|h|^{1/5}$ with $A_{\pm}$ depending on the sign of $h$, as illustrated on Fig.~\ref{k0m1a1b1Cumulant}.

\subsection{Analytical expressions of $p_{\rm end}(x)$ and $p_{\rm ave}(x)$ in $\epsilon \rightarrow 0$}
Finally, we write the explicit analytical expressions of $p_{\rm end}(x)$ and $p_{\rm ave}(x)$ in the $\epsilon \rightarrow 0$ limit.
We consider the biased (unnormalised) probability density $u^{h}$ introduced in the beginning of Section~\ref{sec:fp}. We also consider the same function but with fixed initial condition $u^{h}(x,\tau|x_0,\tau)$. By using these function, we introduced
two logarithmic functions defined as
\begin{equation}
W_{\rm F}(x, t) \equiv \epsilon  \log u^h(x, t), 
\end{equation}
\begin{equation}
W_{\rm B}(x, t) \equiv \epsilon  \log \int  u^{h}(y,t|x,0) \dd y .
\end{equation}
From the generalised Feynman-Kac formula (\ref{supp:eq:18}), we obtain the time evolution equation for them as
%We first define two types of biased distribution function as
%\begin{equation}
%u_{\rm F}(x_\tau,\tau) = \ \left \langle \delta ( x(\tau) - x_\tau) \ee^{h \tau \Lambda(\tau) /\epsilon} \right \rangle_{\rm st},
%\label{ForwardFeynmanKac}
%\end{equation}
%\begin{equation}
%u_{\rm B}(x_0,\tau) = \left \langle \ee^{h \tau \Lambda(\tau) /\epsilon} \right \rangle_{x_0},
%\label{BackwardFeynmanKac}
%\end{equation}
%where the subscript `$x_0$' means that initial condition is $x(0)=x_0$. These two functions satisfy the forward and the backward Feynman-Kac formula (See \cite{Chetrite_Touchette2} or (\ref{supp:eq:18}), for example):
%\begin{equation}
%\frac{\partial }{\partial t} u_{\rm F}(x, t ) = (L_{\rm FP}^{F})[u_{\rm F}] + \frac{h}{\epsilon} \lambda(x) u_{\rm F}(x,  t ),
%\label{Eq.uF}
%\end{equation}
%and
%\begin{equation}
% \frac{\partial }{\partial t} u_{\rm B}(x,  t) = (L_{\rm FP}^{F})^{\dagger}[u_{\rm B}] + \frac{h}{\epsilon} \lambda(x) u_{\rm B}(x,  t ),
%\label{Eq.uB}
%\end{equation}
%where $L_{\rm FP}^{F}$ is the Fokker-Planck operator defined as (\ref{supp:eq:21}) and $(L_{\rm FP}^{F})^{\dagger}$ is its adjoint operator.
%We then use the Cole-Hopf transformation in the Feynman-Kac formula (\ref{ForwardFeynmanKac}) and (\ref{BackwardFeynmanKac}):
\begin{equation}
\begin{split}
&\frac{\partial }{\partial t} W_{\rm F}(x, t) \\
 &= - \epsilon \frac{\partial }{\partial x} F(x) -  F(x) \frac{\partial}{\partial x }  W_{\rm F}(x, t) \\
& \qquad + \epsilon \left (  \frac{\partial }{\partial x } \right ) ^2 W_{\rm F}(x, t)  +  \left ( \frac{\partial }{\partial x}W_{\rm F}(x, t) \right )^2 + h\lambda(x).
\label{WFtimeevolution}
\end{split}
\end{equation} 
and 
\begin{equation}
\begin{split}
&\frac{\partial }{\partial t} W_{\rm B}(x,  t) \\
&=  F(x)   \frac{\partial}{\partial x} W_{\rm B}(x,  t) + \epsilon \left ( \frac{\partial }{\partial x } \right ) ^2 W_{\rm B}(x,t) \\
& \qquad +  \left ( \frac{\partial }{\partial x}W_{\rm B}(x,  t) \right )^2 + h\lambda(x).
\label{WBtimeevolution}
\end{split}
\end{equation} 
These equations can be solved in 
$\epsilon =0$ with $t$ large limit. 
Indeed, by setting $W_{\rm F}(x,t) = t G(h) + W_{\rm F}(x)$ and $W_{\rm B}(x,\tau - t) = (\tau - t) G(h) + W_{\rm B}(x) $ with $G(h) \equiv \lim _{\epsilon \rightarrow 0} G_{\epsilon}(h)$ in these expressions, we obtain the equations to determine $W_{\rm F}(x)$ and $W_{\rm B}(x)$ as
\begin{equation}
\begin{split}
& \frac{\partial W_{\rm F}(x)}{\partial x} =\frac{1}{2} \Bigg [  F(x)  \\
& + C_h(x) \sqrt{F(x)^2 - 4 h \lambda(x) - \min_{y}\left [  F(y)^2 - 4 h \lambda(y)    \right ] }  \Bigg ],
\label{WFfinal}
\end{split}
\end{equation} 
and
\begin{equation}
\begin{split}
& \frac{\partial W_{\rm B}(x)}{\partial x} =\frac{1}{2} \Bigg [ - F(x) \\
& +C_h(x)  \sqrt{F(x)^2 - 4 h \lambda(x) - \min_{y}\left [  F(y)^2 - 4 h \lambda(y)    \right ] }  \Bigg ].
\label{WBfinal}
\end{split}
\end{equation} 
with
\begin{equation}
C_h(x) =  1  \qquad (x<x_{\rm min}),
\end{equation}
\begin{equation}
C_h(x) =  -1 \qquad (x>x_{\rm min}),
\end{equation}
where 
\begin{equation}
x_{\rm min} \equiv {\rm Argmin}_x \left [  F(x)^2 - 4 h \lambda(x)    \right ].
\end{equation}

%\begin{equation}
%\begin{split}
%G(h) &=  -  F(x) \frac{\partial}{\partial x }  W_{\rm F}(x)  +  \left ( \frac{\partial }{\partial x}W_{\rm F}(x) \right )^2 + h\lambda(x),
%\end{split}
%\end{equation} 
%\begin{equation}
%G(h) =  F(x)   \frac{\partial}{\partial x} W_{\rm B}(x)  +  \left ( \frac{\partial }{\partial x}W_{\rm B}(x) \right )^2 + h\lambda(x).
%\end{equation} 
%Since these two equations are quadratic equations with respect to $ \frac{\partial }{\partial x}W_{\rm B}(x)$, and also since $G(h)$ is determined from the variational principle (\ref{steadyassumption}), we thus get 
%\begin{equation}
%\begin{split}
%\frac{\partial W_{\rm F}(x)}{\partial x} =\frac{1}{2} \left [  F(x) \pm \sqrt{F(x)^2 - 4 h \lambda(x) - \min_{y}\left [  F(y)^2 - 4 h \lambda(y)    \right ] }  \right ],
%\label{WFpm}
%\end{split}
%\end{equation} 
%and
%\begin{equation}
%\frac{\partial W_{\rm B}(x)}{\partial x} =\frac{1}{2} \left [ - F(x) \pm \sqrt{F(x)^2 - 4 h \lambda(x) - \min_{y}\left [  F(y)^2 - 4 h \lambda(y)    \right ] }  \right ].
%\label{WBpm}
%\end{equation} 
%For determining the sign, we need to assume that the solution is continuous, from which we find that the sign cannot be changed except for $x=x_{\rm min}$.
%Furthermore, from the assumption that the first derivative can be uniquely defined at $x_{\rm min}$, we find that this sign cannot be the same before and after $x_{\rm min}$. Finally, by assuming
%that these expressions continuously change with respect to the change of $h$ \footnote{From this, we can determine the sign by considering $h=0$.}, we arrive at

Equations (\ref{WFfinal}) and (\ref{WBfinal}) are the key result in this subsection. From them, we indeed get
\begin{equation}
\begin{split}
&p_{\rm end}(x) \sim \exp \Bigg [  (1/\epsilon) \int ^{x} \frac{1}{2} \Big [  F(y)  \\
 &+ C_h(y) \sqrt{F(y)^2 - 4 h \lambda(y) - \min_{z}\left [  F(z)^2 - 4 h \lambda(z)    \right ] }  \Big ] \dd y \Bigg ]
\label{distributionppop}
\end{split}
\end{equation}
and
\begin{equation}
\begin{split}
& p_{\rm ave}(x) \sim \exp \Bigg [  (1/\epsilon) \int ^{x} C_h(y) \\
& \times  \sqrt{F(y)^2 - 4 h \lambda(y) - \min_{z}\left [  F(z)^2 - 4 h \lambda(z)    \right ] } \: \dd y  \Bigg ].
\label{distributionpaux}
\end{split}
\end{equation}
Also from the same equations, we get the most probable $x$ in $p_{\rm end}(x)$ and $p_{\rm ave}(x)$ with $\epsilon \rightarrow 0$.
We denote them by $x_{\rm end}$ and $x_{\rm ave}$, respectively. 
Then, from (\ref{distributionppop}) and (\ref{distributionpaux}), we find that these values satisfy
\begin{equation}
x_{\rm ave} = {\rm Argmax}_{x_{\rm st}} \ G_{\rm st}(x_{\rm st})
\label{xaux_Appendix}
\end{equation}
where $G_{\rm st}(h)$ is defined in (\ref{steadyassumption}), and
\begin{equation}
\frac{F(x_{\rm ave})^2}{4h} =  \lambda(x_{\rm ave}) - \lambda(x_{\rm end}). 
\label{xauxapop_Appendix}
\end{equation}
Since $\frac{F(x_{\rm ave})^2}{4h}\neq 0$, $x_{\rm ave}$ and $x_{\rm end}$ are different from each other. In other words,  $p_{\rm ave}$ and $p_{\rm end}$ concentrate on different values of their argument in the $\epsilon \rightarrow \infty$ limit, as announced in the main text.

\begin{figure*}[h]
\centering
%\subfigure[$\partial W_{\rm F}(x) / \partial x$ for $h=1$]{
\includegraphics[width=0.9\columnwidth]{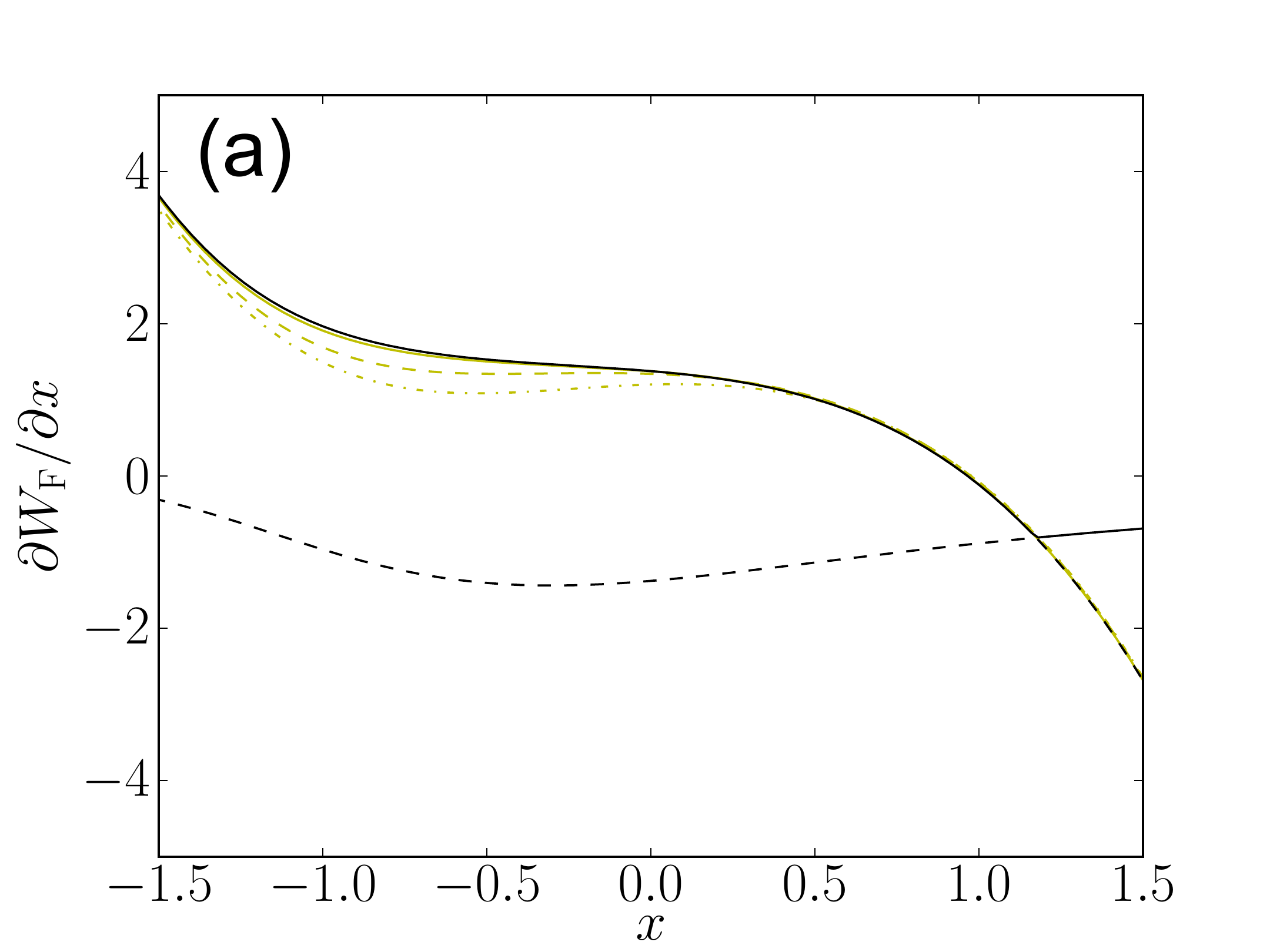}%}
%\subfigure[$\partial W_{\rm F}(x) / \partial x$ for $h=-1$]{
\includegraphics[width=0.9\columnwidth]{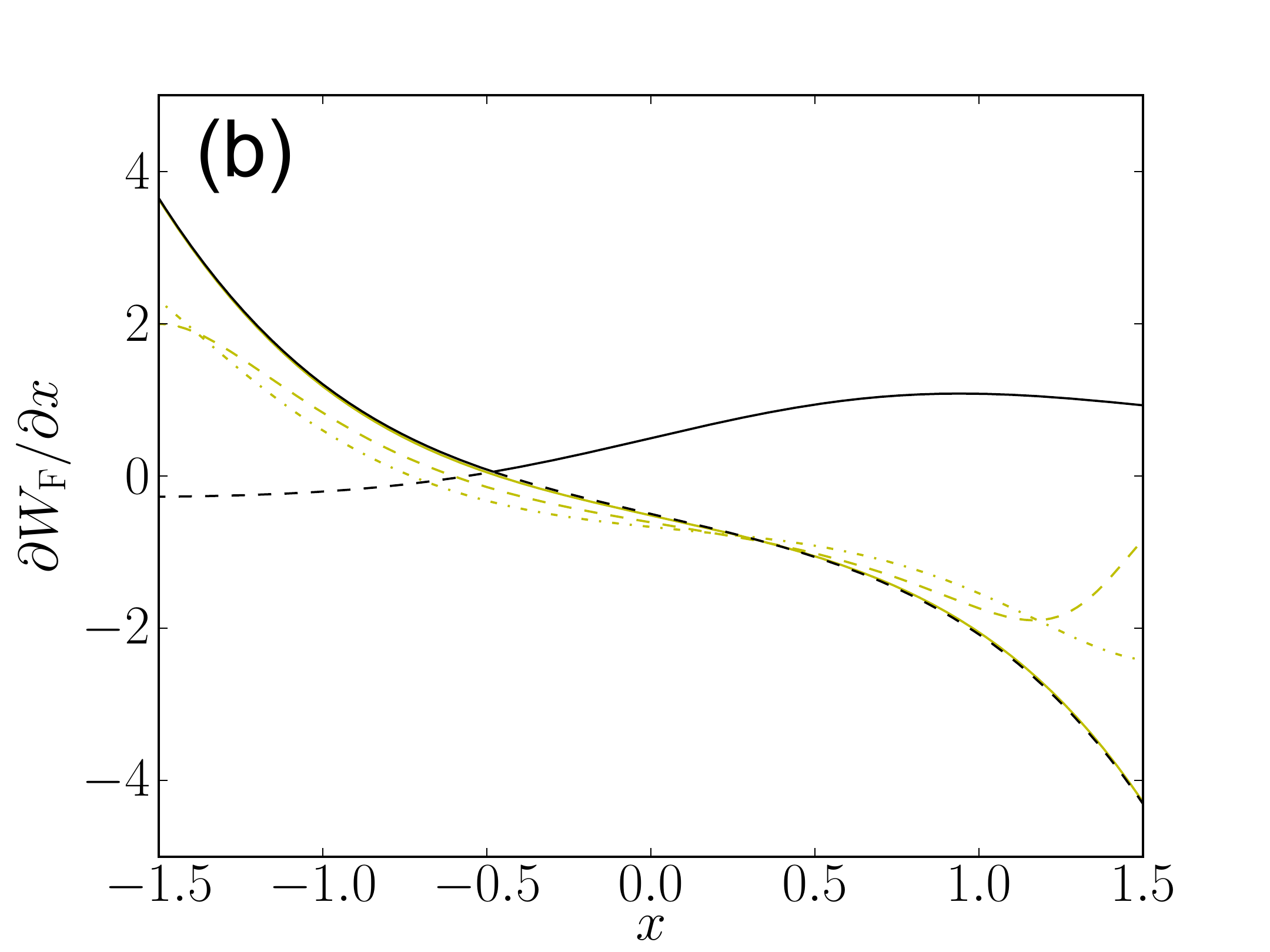}\\%}
%\subfigure[$\partial W_{\rm B}(x) / \partial x$ for $h=1$]{
\includegraphics[width=0.9\columnwidth]{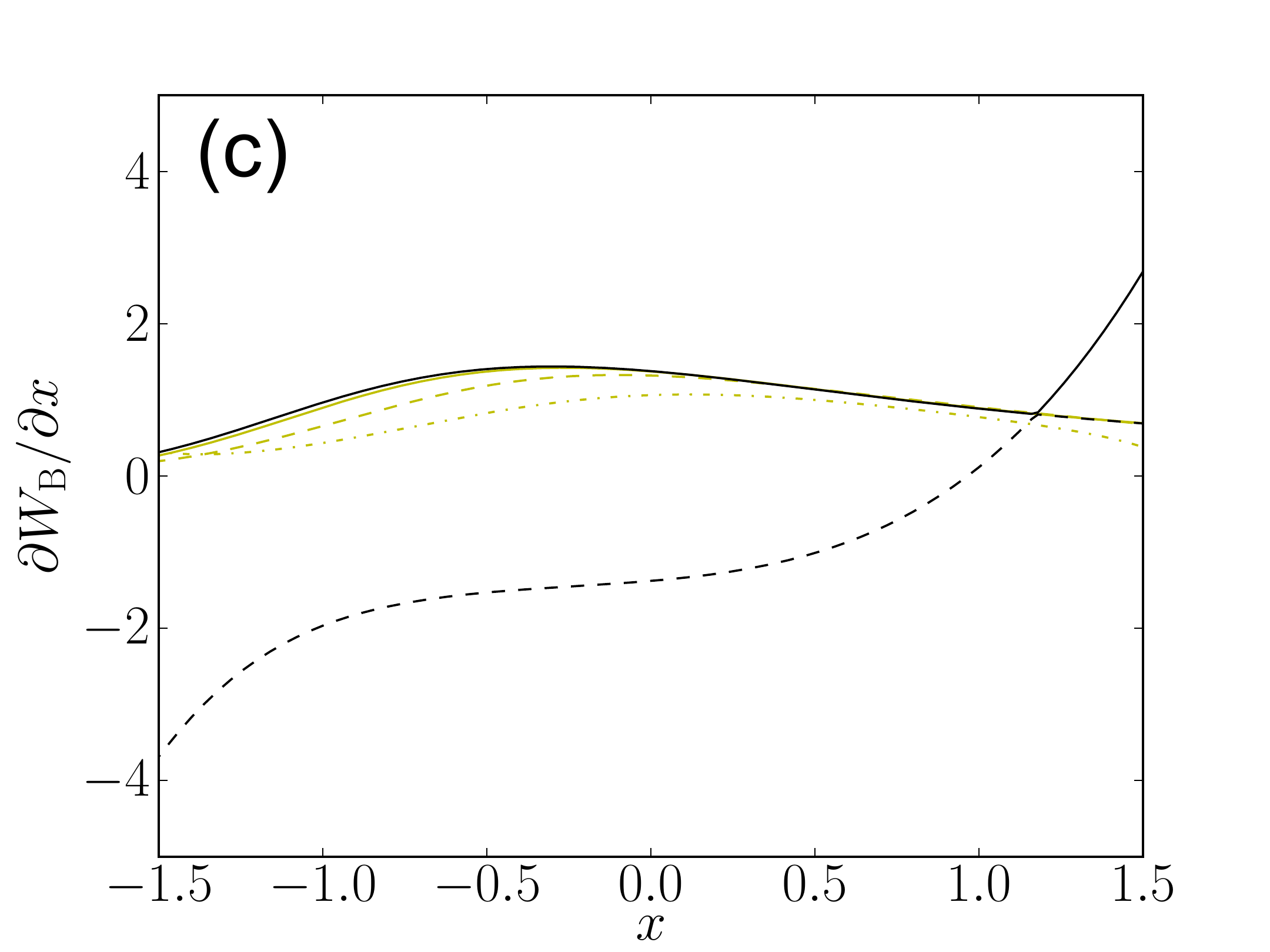}%}
%\subfigure[$\partial W_{\rm B}(x) / \partial x$ for $h=-1$]{
\includegraphics[width=0.9\columnwidth]{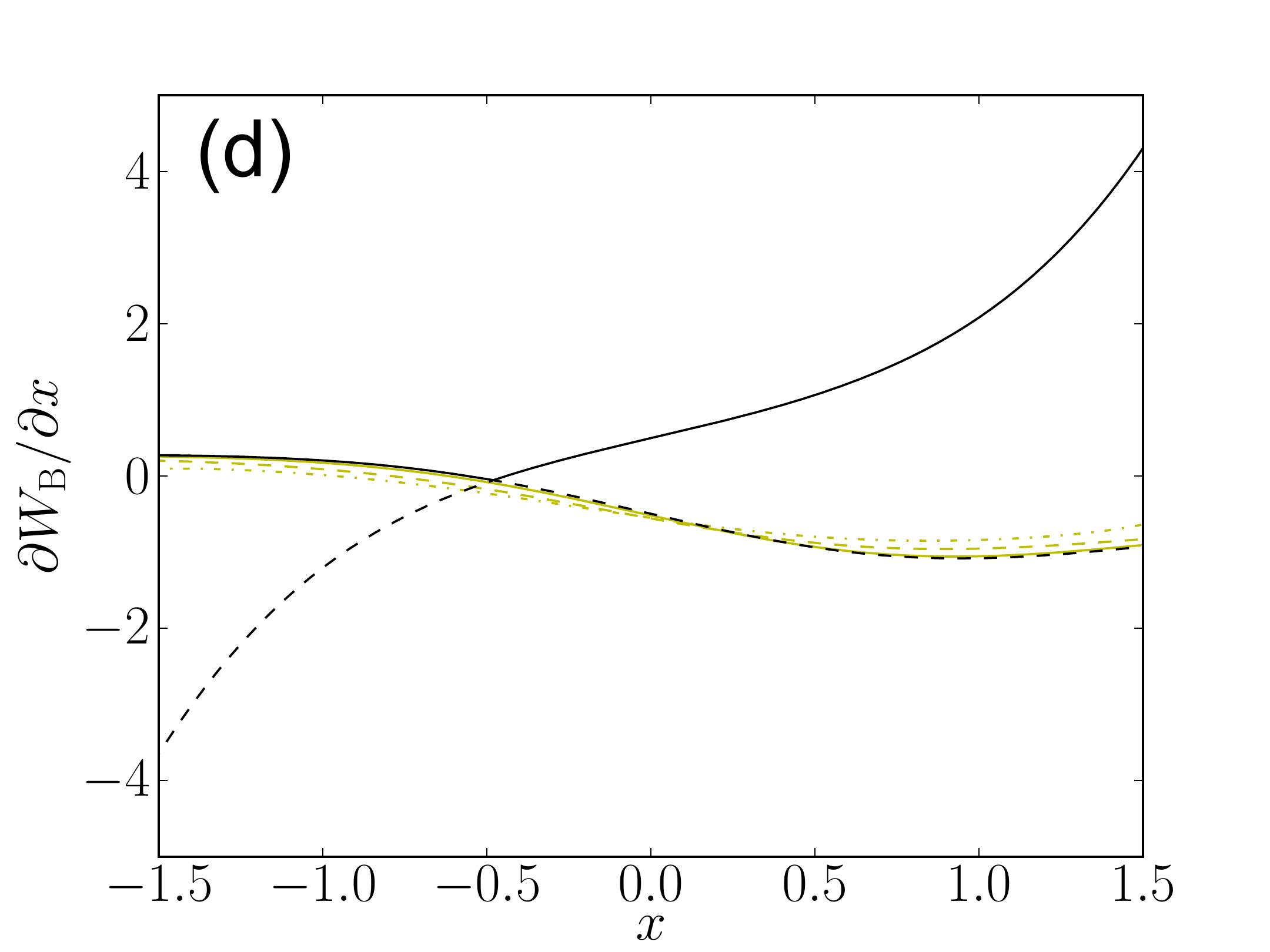}%}
\caption{\label{Forwardh1} 
(Color online) The functions $\partial W_{\rm F}(x,t) / \partial x$ ((a) and (b)) and $\partial W_{\rm B}(x,t) / \partial x$ ((c) and (d))
obtained in the large $t$ limit
by solving numerically~\eqref{WFtimeevolution} and~\eqref{WBtimeevolution} [yellow lines or light grey lines in the printed version].
We set $h=1$ ((a) and (c)) and $h=-1$ ((b) and (d)).
The line types correspond different values of $\epsilon$: dash-dotted, dashed and solid lines correspond to $\epsilon =1,0.5,0.1$, respectively. To illustrate the determination of the $\pm$ sign of $C_h$ in the analytical results (\ref{WFfinal}) and (\ref{WBfinal}), we also plot on each subfigure
those results with the choice of $C_h=1$ (for all $x$) as black solid lines and the choice of  $C_h=-1$ (for all $x$) as black dashed lines.
As the noise goes to zero, we observe the convergence of the functions $\partial W_{\rm F}(x,t) / \partial x$ and $\partial W_{\rm B}(x,t) / \partial x$ determined numerically  at large $t$ towards the analytical line (\ref{WFfinal}) and (\ref{WBfinal}), where the  $+$ sign in $\pm$ is taken for $x<x_{\rm min}$ and the $-$ sign is taken for $x>x_{\rm min}$.}
\end{figure*}

For checking the validity of the obtained expressions, we numerically solve the equations 
(\ref{WFtimeevolution}) and (\ref{WBtimeevolution}) during a sufficiently large time interval $t$.
We set $h=1$ (Fig.~\ref{Forwardh1}(a) and Fig.~\ref{Forwardh1}(c)) and $h=-1$ (Fig.~\ref{Forwardh1}(b) and Fig.~\ref{Forwardh1}(d)).
The different colours represent the different values of $\epsilon$: yellow, blue, red lines correspond to $\epsilon =1,0.5,0.1$, respectively. In the same figure, we plot 
the analytical lines (\ref{WFfinal}) and (\ref{WBfinal}), with $C_h=1$ (for all $x$) (black solid line) and $C_h=-1$ (for all $x$) (black dashed line).
We can see the convergence of the numerical lines (with decreasing $\epsilon$) towards the analytical lines (\ref{WFfinal}) and (\ref{WBfinal}), where  $+$ sign is chosen for $x<x_{\rm min}$ and $-$ sign is chosen for $x>x_{\rm min}$.

\begin{thebibliography}{99}
\bibitem{footnoteinteraction}
All the results presented here can be generalised
to models with $n$ interacting particles, by replacing $d\to nd$ and including interaction forces in $F(x)$.  Markov jump
processes are also workable.
\bibitem{FootnoteConvention_Multiplication}
More precisely, when the Langevin equation (\ref{equ:dx}) and the time-averaged quantity  (\ref{Observe_Vector}) are written in Stratonovich convention, one first transforms these equations to ones in It\=o by using a transformation formula \cite{Gardiner}, and then re-define $F$ and $\Lambda_{\rm d}$ so that these transformed equations have the same forms as (\ref{equ:dx}), (\ref{Observe_general}), (\ref{Observe_Scalar}) and (\ref{Observe_Vector}) (and can be analysed by the same method that is discussed throughout this paper).
\bibitem{Footnote_Work}
It is also related to quantities in equilibrium thermodynamics. Indeed, when we set $\lambda_c$ to be an external force, $\tau \Lambda_{\rm c}(\tau)$ is the work done by the external system.
\bibitem{footnoteinvertible}
The matrix $\kappa$ is assumed to be invertible but the generalization to non-invertible $\kappa$ is direct.
\bibitem{footnote:KullbackLeibler}
$D_2$ is also called the Kullback--Leibler divergence~\cite{KBd}. It is positive and measures the distance between the distributions $p_{\rm ave}$ and $p_{\rm end}$. It is zero if and only if those distributions are equal almost everywhere.
\bibitem{footnotecitation_anneaing}
The numerical calculations exploit an ``annealing procedure'' in which
$\epsilon$ is reduced in steps of size
$\delta\epsilon=0.01$, with the best estimate of $w^*$ from the each step being used as the first guess for $w^*$ 
in the subsequent step.
\bibitem{footnote:m2andD2}
These integrands, which we denote by $\tilde m_2(x)$ and $\tilde D_2(x)$, are proportional to each other when $\delta \equiv |p_{\rm ave}(x) - p_{\rm end}(x)|$ is small. Indeed, these satisfy $\tilde m_2(x) = 2 \tilde D_2(x) + {\mathcal O}(\delta ^3)$. This is the reason why Fig.~\ref{fig:multiplicitiesDiscussions}(a) and Fig.~\ref{fig:multiplicitiesDiscussions}(b) show similarities.
\bibitem{footnote:linearstabilityanalysis}
We can prove this fact from a linear stability analysis. See \cite{InPreparation} for the detail.
\end{thebibliography}
\end{document}